\documentclass[fleqn,usenatbib]{mnras}

\usepackage[T1]{fontenc}
\usepackage{ae,aecompl}

\usepackage{graphicx}	
\usepackage{amsmath}	
\usepackage{amssymb}	
\usepackage[shortlabels]{enumitem}
\usepackage{lineno} 
\usepackage{multirow}
\usepackage[table]{xcolor}

\usepackage{newtxtext,newtxmath}

\usepackage{pdflscape}

\usepackage{float}
\defcitealias{salt3}{K21}
\defcitealias{jla}{JLA}

\title[SALT2 Versus SALT3]{SALT2 Versus SALT3: Updated Model Surfaces and Their Impacts on Type Ia Supernova Cosmology}

\author[G. Taylor et al.]{
G. Taylor$^{1}$\thanks{E-mail: gtaylor@mso.anu.edu.au},
D. O. Jones$^{2,3}$,
B. Popovic$^{4}$,
M. Vincenzi$^{4}$,
R. Kessler$^{5,6}$,
D. Scolnic$^{4}$,
M. Dai$^{7}$,
\newauthor
W. D. Kenworthy$^{7,8}$, 
J. D. R. Pierel$^{9}$
\\
$^{1}$Research School of Astronomy and Astrophysics, Australian National University, Canberra, Australia.\\
$^{2}$Department of Astronomy and Astrophysics, University of California, Santa Cruz, CA 95064, USA.\\
$^{3}$NASA Einstein Fellow.\\
$^{4}$Department of Physics, Duke University, Durham, NC, 27708, USA.\\
$^{7}$Department of Physics and Astronomy, The Johns Hopkins University, Baltimore, MD 21218, USA.\\
$^{8}$The Oskar Klein Centre for Cosmoparticle Physics, Department of Physics,Stockholm University, SE-10691 Stockholm, Sweden.\\
$^{5}$Department of Astronomy and Astrophysics, The University of Chicago, 5640 South Ellis Ave., Chicago, IL 60637, USA.\\
$^{6}$Kavli Institute for Cosmological Physics, University of Chicago, Chicago, IL 60637, USA.\\
$^{9}$Space Telescope Science Institute, Baltimore, MD 21218, USA.
}

\date{Accepted XXX. Received YYY; in original form ZZZ}

\pubyear{2022}

\begin{document}
\label{firstpage}
\pagerange{\pageref{firstpage}--\pageref{lastpage}}
\maketitle

\begin{abstract}
For the past decade, SALT2 has been the most common model used to fit Type Ia supernova (SN\,Ia) light curves for dark energy analyses. Recently, the SALT3 model was released, which upgraded a number of model features but has not yet been used for measurements of dark energy. Here, we evaluate the impact of switching from SALT2 to SALT3 for a SN cosmology analysis.
We train SALT2 and SALT3 on an identical training sample of 1083 well-calibrated Type Ia supernovae, ensuring that any differences found come from the underlying model framework. We publicly release the results of this training (the SALT "surfaces"). We then run a cosmology analysis on the public Dark Energy Survey 3-Year Supernova data sample (DES-SN3YR), and on realistic simulations of those data. We provide the first estimate of the SN+CMB systematic uncertainty arising from the choice of SALT model framework (i.e. SALT2 versus SALT3), $\Delta w$ = +0.001 $\pm$ 0.005 --- a negligible effect at the current level of dark energy analyses. We also find that the updated surfaces are less sensitive to photometric calibration uncertainties than previous SALT2 surfaces, with the average spectral energy density dispersion reduced by a factor of two over optical wavelengths. This offers an opportunity to reduce the contribution of calibration errors to SN cosmology uncertainty budgets.

\end{abstract}

\begin{keywords}
dark energy -- software: data analysis -- transients:  supernovae -- supernovae: general 
\end{keywords}



\section{Introduction}

Type Ia supernova explosions (SNe~Ia) are luminous enough to be observed at great distances ($z \approx 2$, \citealt{Jones_2013_highz, Rodney_2014_highzSN, Graur_2014_highzSN, Riess_2018_highzSN}), and consistent enough to be used as standardisable candles for measuring such distances. For the past two decades, they have been a key cosmological probe in the discovery and quantification of dark energy. Reviews of SNe~Ia usage in cosmology can be found in e.g., \citet{howell_11, cosmic_accln_review, dansreview}.

As dedicated surveys have added to the total sample of observed SNe~Ia, cosmology samples have grown in statistical power from the first analyses of $\approx$ 40 SNe~Ia \citep{riess98, perlmutter99}, to rolling surveys with hundreds to thousands of SNe~Ia (e.g., \citealt{jla, panstarrs18, Abbott_2019, panplus_constraints}), to the next generation of surveys which anticipate samples of $10^4$ - $10^5$ SNe~Ia \citep{lsst_2019, roman_21}. The focus is now on improving the methods used in SN~Ia cosmology to take full advantage of the forthcoming improved statistics. One area of potential improvement is the model used for light curve fitting --- a crucial step in obtaining the SN parameters needed to standardise the luminosity and measure distances.

The majority of modern analyses have made use of the SALT model framework \citep{salt} for light curve fitting \citep{g10_scatter, conley11, jla, sdssII, des3yr_systematics, panstarrs18, Riess_2018, Jones_2019, panplus_constraints}. Alternative light curve models (e.g., SNooPy, \citealt{snoopy}) exist but are not discussed here, as they are not typically used for measuring the dark energy equation-of-state parameter $w$;\footnote{Comparisons of different light curve models can be found in the literature (e.g., \citealt{compare, kessler09}).} the ubiquity of the SALT framework indicates that it will be used in future analyses and therefore must be maintained and updated where appropriate. Such updates are the focus of this paper. 

\begin{figure*}
  \begin{center}
   \includegraphics[width=1.5\columnwidth]{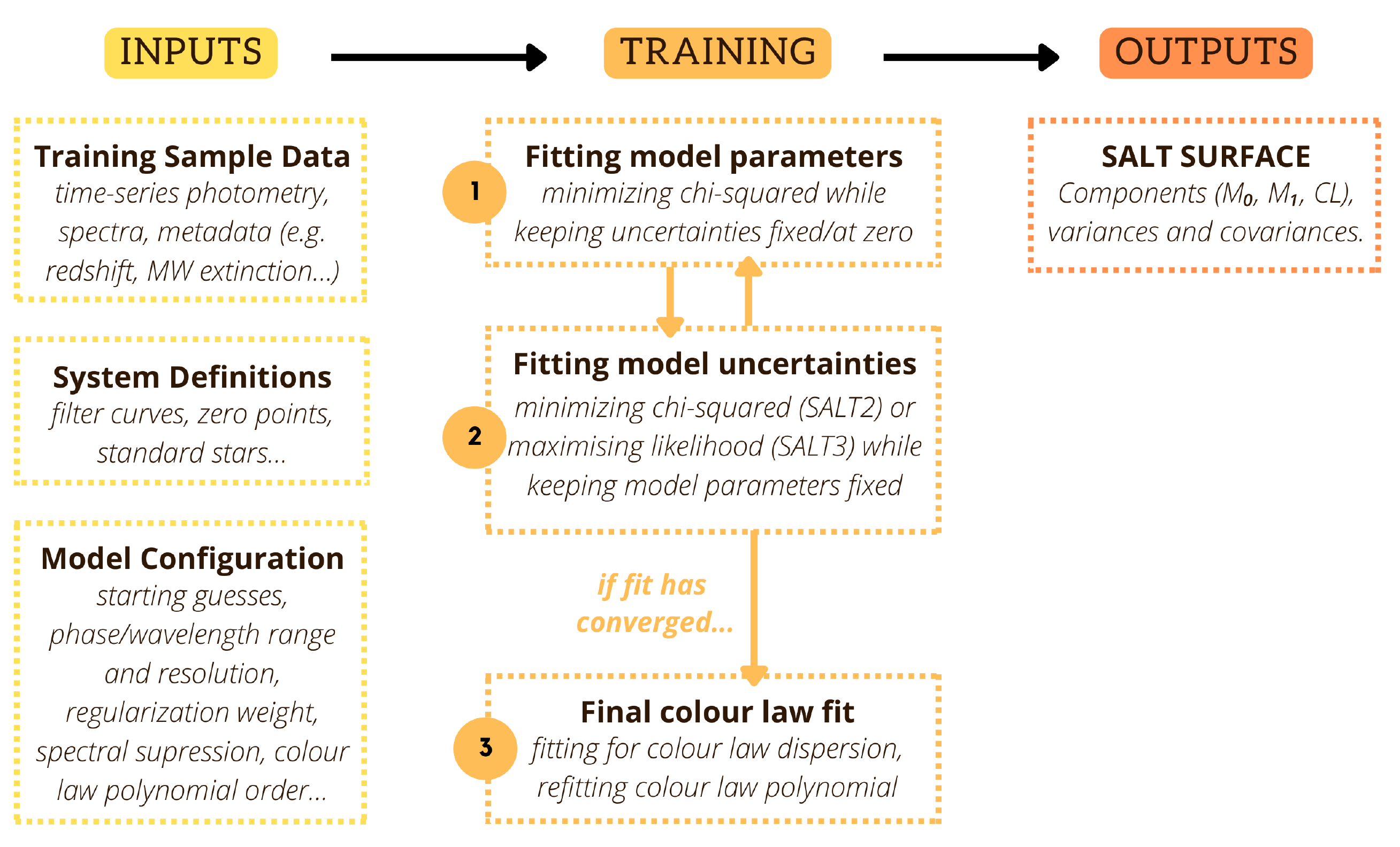}
  \end{center}
  \caption{General overview of the SALT training procedure.}
  \label{training_diagram}
\end{figure*}

SALT2 \citep{salt2} is the successor of SALT \citep{salt}; the SALT2 training output (i.e. surface) that is most widely used for cosmology was released in \citet[hereafter \citetalias{jla}\footnote{JLA = Joint Light Curve Analysis between SDSS and SNLS teams.}]{jla}, though two more recent  surfaces have also been published \citep{T21, superfrag}. Recently, the SALT3 model was released \citep[hereafter \citetalias{salt3}]{salt3}, including a SALT3 surface which was trained with the open source \verb|SALTShaker|\footnote{\url{https://saltshaker.readthedocs.io}} Python package. \verb|SALTShaker| closely follows the procedure of the original SALT2 training code (\verb|snpca|), with minor differences in implementation described in \citetalias{salt3} and summarised in this work. 

The accessibility of \verb|SALTShaker| points towards SALT3 being adopted in the next generation of SN cosmology analyses. For example, \citet{salt3nir} presents a SALT3 model extended to cover near-infrared wavelengths, a critical step needed to leverage the full power of SN~Ia data from the upcoming \textit{Nancy Grace Roman Space Telescope}. \citet{jones_22_host} used SALTShaker to explore the effect of including the relationship between SNe Ia and their host-galaxy properties in the SALT3 model surfaces. However, the SALT3 model has not yet been used in a SN cosmology analysis.

In this paper, we detail the known similarities and differences in the SALT2 and SALT3 model frameworks ($\S$~\ref{section:SALT2-training}), highlighting any deviations in the modelling that could impact a cosmology analysis. We present two new SALT surfaces: SALT2.FRAG and SALT3.FRAG; these are named in reference to the Supercal-Fragilistic re-calibration from \citealt{superfrag}, which is currently the most modern and expansive cross-calibration of photometric systems used to observe SN~Ia cosmology samples. We also present associated ``systematic uncertainty surfaces" that are needed to evaluate systematic uncertainties in a 
SN~Ia cosmology analysis: sets of SALT surfaces where the input photometric calibration has been varied slightly. (\S~\ref{frag_surfaces}). These surfaces are trained on the same \citetalias{salt3} data set of 1083 SNe~Ia, using the same advanced photometric calibration \citep{superfrag}, with the \verb|snpca| and \verb|SALTShaker| training schemes respectively. Differences between SALT2.FRAG and SALT3.FRAG will therefore be a result of the different training processes/model frameworks, rather than different training inputs. 

We test the performance of the new surfaces by running a SN cosmology analysis with each, based on the Dark Energy Survey's 3-year Supernova Analysis (DES-SN3YR; \citealt{Abbott_2019, des3yr_systematics}). We use both real DES-SN3YR data and simulations to assess which updated SALT surface is better at recovering the SN parameters and moreover, distances and cosmological parameters\ ($\S$~\ref{section:analysis}). We discuss our findings, and make recommendations about the treatment of SALT surfaces in upcoming analyses ($\S$~\ref{sec:discussion}).

\section{SALT Model Framework}
\label{section:SALT2-training}

\subsection{General Similarities Between SALT2 and SALT3}

The empirical spectro-photometric SALT model is defined similarly for SALT2 and SALT3. The spectral flux density $F$ for each SN~Ia at a particular phase ($p$) and wavelength ($\lambda$) is 
\begin{linenomath*}
  \begin{align} 
    \label{flux1}
    \begin{split}
    F(p, \lambda) = x_{0} \times\left[M_{0}(p, \lambda)+x_{1} M_{1}(p, \lambda)\right] \times \exp [c C L(\lambda)].
    \end{split}
  \end{align}
\end{linenomath*}

\noindent
The SALT model components are $M_0$ (spectral time-series describing the mean spectral energy distribution (SED) of a SN~Ia, $x_1=0, c=0$), $M_1$ (spectral time-series describing the first-order deviation around $M_0$, which are correlated with light curve width variations), and $CL$ (wavelength-dependent colour law capturing the phase-independent colour variation from combined intrinsic SN colour and host-galaxy dust extinction). These components are determined from the model training process. We name this collection of trained components (along with their associated variances and covariances to estimate un-modeled variability) a SALT "surface".

The SALT2 and SALT3 training processes follow the same general framework, shown in Figure~\ref{training_diagram}. We note that in previous releases of SALT2, the surfaces included the colour law from step 1 of the training process
instead of the final colour law in step 3 (as prescribed in e.g. \citealt{mosher14}). In this work, we use the final SALT2 colour law. More details about the impact of this update on previously released SALT2 surfaces are given in Appendix~\ref{app:cl}.

A SALT surface can be used to fit light curves of an individual SN from time-series photometry to determine its parameters: $x_0$ (overall flux normalization), $x_1$ (strongly correlated with stretch), and $c$ (colour). These fitted parameters can be used to calculate the distance modulus from the (modified) Tripp equation \citep{tripp98}:

\begin{linenomath*}
  \begin{align} 
    \label{mod_tripp}
    \mu=m_B-M+\alpha x_1-\beta c-\delta \mu_{\text {host }}-\delta \mu_{\text {bias }}
  \end{align}
\end{linenomath*}

\noindent where $m_B =-2.5\log(x_0)$, $M$ represents the absolute magnitude of a SN~Ia with $c = x_1 = 0$, $\delta \mu_{\rm host}$ is a mass-step correction, $\delta \mu_{\rm bias}$ is a bias-correction term to account for selection effects, and $\alpha$ and $\beta$ are nuisance parameters representing the slopes of the stretch-luminosity and colour–luminosity relations.

\subsection{Differences Between SALT2 and SALT3}
\label{section:23_differences}

\subsubsection*{Accessibility of Training Programs}

SALT2 is trained with \verb|snpca|, a
FORTRAN/C++ program based on principle component analysis. A detailed description of this training process is provided in \citet{g10_scatter} and \citet{mosher14}, though the training code itself is not publicly available. 

SALT3 is trained with the open-source \verb|SALTShaker| Python program, making it much more accessible and easy to update than the SALT2 training procedure. SALT3 is also able to be trained with data in native \verb|SNANA| format (a widely-used SN analysis package, \citealt{snana09}), whereas SALT2 requires the data to be converted to its own \verb|snfit| format (which is not widely used). A detailed description of the SALT3 training process is provided in \citetalias{salt3}.

\subsubsection*{Model Ranges and Resolutions}

The SALT2 model is nominally defined over 2000-9200\AA\ and -20 to +50 days from $B$-band peak magnitude. Using additional data for training (e.g., $z$-band observations from Foundation, \citealt{foundation_dr1}), SALT3 is defined over 2000-11000\AA\ and -20 to +50 days days\footnote{While in theory the phase and wavelength ranges of either model can be extended by modifying the training inputs, the design of \texttt{snpca} makes this a non-trivial task for SALT2.}.

There are some regions of rest-frame wavelength space where the model may be defined and trained, but it is not used for light curve fitting because of large model uncertainties. These uncertainties depend on the quality of the training sample used; because SALT3 includes $\lambda \geq 9200$\AA\ data in its training process, the resulting surface can be used to fit light curve data at longer wavelengths than SALT2. This is discussed further in $\S$~\ref{methods}.

In both SALT models, the $M_0$ and $M_1$ spectral time series are constructed with third-order b-spline basis functions (resembling Gaussians). The amplitude of each basis function is allowed to vary during the training.

To allow higher resolution in some areas of the parameter space, the locations of spline basis knots are not uniformly distributed.
SALT2 uses 14 phase basis functions (spanning an average of $\approx5$ days) and 100 wavelength basis functions (spanning an average of $\approx72$\AA). SALT3 uses 20 phase basis functions (spanning an average of $\approx3$ days) and 127 wavelength basis functions (spanning an average of $\approx$72\AA).

\subsubsection*{Colour Laws}

The SALT colour law ($CL$) is defined such that the flux-change for a given SN with colour parameter $c$ at a particular wavelength is $\exp\left[cCL(\lambda)\right]$ (Equation~\ref{flux1}). In defining the colour law, one first defines the \textit{reduced} wavelength, $\lambda_r(\lambda) = \frac{\lambda - \lambda_B}{\lambda_V - \lambda_B}$, where $\lambda_B$ and $\lambda_V$ are the central wavelengths of the $B$- and $V$-band filters
. To define $CL(\lambda)$,

$$
\begin{aligned}
C L_p(\lambda) &= \alpha \lambda_r + \sum_{i=1}^{N_{C L}} \boldsymbol{cl}_i \lambda_r^{i+2}, ~\rm where~\alpha = 1 - \sum_{i=1}^{N_{C L}}  \boldsymbol{cl}_i\\
C L(\lambda) &=\left\{\begin{array}{ll}
C L_p^{\prime}\left(\lambda_{-}\right)\left(\lambda-\lambda_{-}\right)+C L_p\left(\lambda_{-}\right) & \lambda<\lambda_{-} \\
C L_p^{\prime}\left(\lambda_{+}\right)\left(\lambda-\lambda_{+}\right)+C L_p\left(\lambda_{+}\right) & \lambda>\lambda_{+} \\
C L_p(\lambda) & \text { otherwise }
\end{array}\right.
\end{aligned}
$$

\noindent where $C L_p^{\prime}$ is the derivative with respect to $\lambda$ and $cl_i$ are the fitted polynomial coefficients.

The $CL$ input parameters for SALT2 (fifth-order polynomial) and SALT3 (sixth-order polynomial) are shown in Table~\ref{cl_table}. The colour law is linearly extrapolated from the polynomial endpoints ($\lambda_-, \lambda_+$) to the model endpoints for both SALT2 and SALT3.

\begin{table}
\renewcommand\thempfootnote{\arabic{mpfootnote}}
\begin{minipage}{0.4\textwidth}
\renewcommand{\arraystretch}{1.5}
\centering
\caption{Colour law input parameters for SALT2 and SALT3 models.}
\label{cl_table}
\begin{tabular}{llllll} \textbf{Model}
 & \textbf{$N_{CL}$} & \textbf{$\lambda_-$} & \textbf{$\lambda_+$} & \textbf{$\lambda_B$} & \textbf{$\lambda_V$} \\ \hline
SALT2 & 4 & 2800\AA & 7000\AA & 4302.57\AA & 5428.55\AA \\ \hline
SALT3 & 5 & 2800\AA & 8000\AA & 4302.57\AA & 5428.55\AA \\ \hline
\end{tabular}
\end{minipage}
\end{table}

\subsubsection*{Additional Model Definitions}

Definitions used to constrain both models are described in Appendix~\ref{common_model_definitions}, and here we describe model definitions that differ between the two models.

SALT2 requires that the rest-frame synthetic $V$-band flux of the $M_1$ component at the time of $M_0$ peak magnitude is defined to be 0, implying that $x_1$ has no effect on observed $B-V$ colour at this peak phase. SALT3 does not apply this definition, instead requiring that the distributions of $x_1$ and $c$ have no correlation in the training sample. This SALT3 constraint implies that the observed colour, $c$ (which includes host galaxy dust contributions) is independent of the processes that govern stretch, $x_1$, at any phase. In practise, the two approaches are very similar \citepalias{salt3}.

\subsubsection*{Uncertainty in the SALT Surface}

Empirical models of SNe~Ia, such as SALT, do not fully describe the observed variability of the SNe~Ia population, leaving some remaining un-modelled "intrinsic scatter".\footnote{\citet{Rubin_2020_twins} and \citet{2020Rose} suggest that three to five parameters drive SN~Ia SED variation; SALT has a single parameter ($x_1$) with which to model the entire variation in the SNe~Ia population. Both SALT2 and SALT3 are capable of extending their parameterization as more training data becomes available.}
This uncertainty in the SALT model, referred to as in-sample variance, is treated as noise independent from one broadband to another. Correlated uncertainty within bands is accounted for by an intrinsic colour scatter component, which assigns constant covariance for all observations in a given band (and zero covariance between points in different bands). In-sample variance parameters are included in the training, along with an out-of-sample variance that estimates the uncertainty due to the finite training sample size and signal-to-noise ratio (SNR).

SALT2 assumes that the in-sample variance has the same form as the out-of-sample variance, simply scaling the latter by a smooth "error snake" that was fit during the training. SALT2 reports \textit{relative} component uncertainties that must be scaled by the model flux during the light curve fitting process.

SALT3, on the other hand, iteratively estimates the in-sample variance during the model training procedure.  This is performed by maximizing the log-likelihood of the model while fitting the in-sample variance parameters, defined by a series of zeroth-order b-spline functions (equivalent to nearest-neighbor interpolation), and keeping the model parameters fixed (training step 2 in Figure~\ref{training_diagram}).  The out-of-sample variance is computed from the inverse Hessian matrix at the conclusion of the training process, and the \textit{absolute} uncertainty (in units of flux) is reported for each component. This is described in detail in $\S$~2 of \citetalias{salt3}.

\begin{figure*}
\centering
\begin{minipage}[t]{.48\textwidth}
\begin{center}
   \includegraphics[width=\columnwidth]{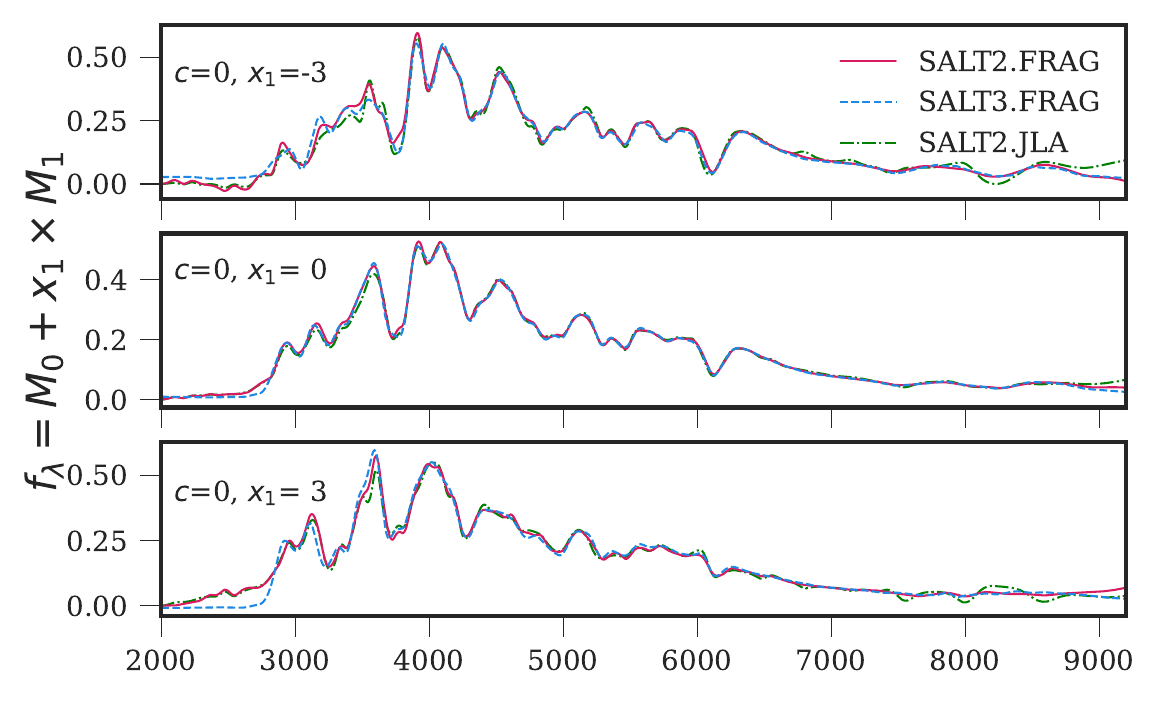}
  \end{center}
  \caption{The trained SED of $c=0$ SN~Ia (at phase~=~0.0) for the SALT2.FRAG \textit{(pink, solid)} and SALT3.FRAG \textit{(blue, dashed)} surfaces. For comparison, the published SALT2.JLA \textit{green, dash-dotted} surface is also shown. The panels show the SED for $x_1 = -3, 0, +3$ respectively.}
  \label{m0-m1}
\end{minipage}\qquad
\begin{minipage}[t]{.48\textwidth}
\begin{center}
   \includegraphics[width=\columnwidth]{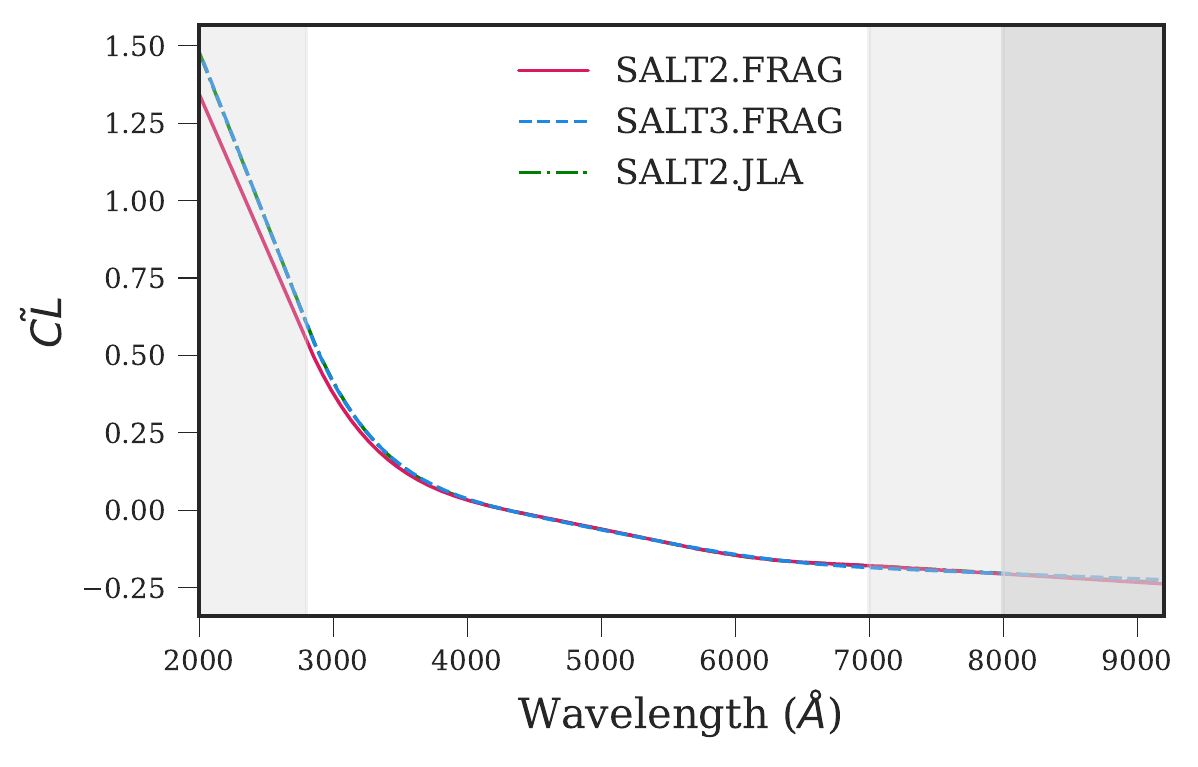}
  \end{center}
  \caption{For $c=-0.1$, $cCL(\lambda)$ is shown for SALT2.FRAG \textit{(pink, solid)} and SALT3.FRAG \textit{(blue, dashed)} surfaces. The published SALT2.JLA colour law \textit{(green, dash-dotted)} is also shown; this overlaps the SALT3.FRAG colour law for most of the wavelength range. The colour laws of both models are linearly extrapolated for wavelengths shorter than 2800\AA\ and longer than 7000\AA (for SALT2, light-grey shaded region) and 8000\AA (for SALT3, dark grey shaded region).}
  \label{cL}
\end{minipage}
\end{figure*}

\subsubsection*{Regularization}

Regularization adjusts the "smoothness" of the SALT model over regions with limited or missing spectral data by penalizing model fits with ringing artefacts.\footnote{Ringing is any unphysical rapid fluctuation in the SED, caused by the fitting with only broadband photometry in regions without spectra.} However, it is important not to smooth out real spectral features, which sometimes resemble ringing artefacts.

SALT2 applies wave-gradient and dyadic regularization. Gradient regularization penalizes changes in flux with respect to wavelength, and ignores changes with respect to phase. Dyadic regularization penalizes fluxes that cannot be decomposed into separate functions of phase and wavelength. SALT2 applies no regularization to penalize ringing that appears only in phase-space.

SALT3 implements a different regularization scheme, which includes phase gradient (penalizes high-frequency flux changes with respect to phase), wave gradient, and dyadic regularization. Due to differences in application, the regularization weights between SALT2 and SALT3 cannot be directly compared - but both models only apply regularization in regions where the data density is sufficiently low.

\section{SALT Surfaces presented in this paper}
\label{frag_surfaces}

\subsection{K21+ Fragilistic Training Sample}
\label{section:tsample}

Here we describe the ''K21+" training sample used to produce the SALT surfaces presented in this paper. 
We use the data in Table 4 of \citetalias{salt3}, with a few updates described below. The sample contains 1083 SNe~Ia (0.001 $\leq z \leq$ 0.850) with photometric data, 380 of which also have one or more epoch of spectral data (1207 spectra in total).

The heliocentric SN redshifts have been updated in-line with \citet{carr_21}; the mean change in $z$ in the K21+ sample is +0.0008 (the mean \textit{absolute} $z$ change is 0.003), and the mean change in $\sigma_z$ is -0.0005. The CSP photometry has been updated from DR2 \citep{csp_2} to DR3 \citep{CSP_DR3} (affecting 24 SNe~Ia in K21+).

As in \citet{T21} and \citetalias{salt3}, the training sample uses the recalibrated \citet{schlegel98} dust map prescribed in \citet{schlafly2010} and \citet{schlafink}. 

Additionally, the calibration of the photometric systems (used for both the model training and light curve fitting) has been updated based on SuperCal-Fragilistic \citep{superfrag}. This allows the surfaces trained with the K21+ sample (e.g., SALT2.FRAG, SALT3.FRAG) to be used for analyses that calibrate their cosmology samples with the SuperCal-Fragilistic scheme, without introducing a bias from inconsistent calibrations. 

We note that some of the historic SNe zero points were unable to be re-calibrated, as their surveys observed low numbers of stars. To account for this, a 20 mmag calibration uncertainty is adopted (about $3\times$ the typical reported uncertainties by other surveys). This uncertainty is not used in the training of the fiducial SALT.FRAG surfaces, but is used to create the systematic uncertainty surfaces that measure the impact of photometric calibration \textit{on} the fiducial surfaces ($\S$~\ref{subsec:systsurf}). The $u$- and $U$-band filters have also not been re-calibrated, due to lack of overlap with PS1 \citep{ps1} wavelengths.

\subsection{SALT2.FRAG and SALT3.FRAG}

SALT2.FRAG is trained on the K21+ training sample described above, using Supercal-Fragilistic-prescribed instruments and magnitude systems. The model configuration is consistent with that used in \citetalias{jla} and \citet{T21}, with the exception of the applied colour law update (described in $\S$\ref{section:SALT2-training}). SALT3.FRAG is also trained on K21+ and uses the Supercal-Fragilistic-prescribed instruments and magnitude systems. The model configuration is consistent with that used in \citetalias{salt3}.

SALT2.FRAG and SALT3.FRAG are both publicly available online.\footnote{\url{10.5281/zenodo.4001177}} The components of the SALT2.FRAG and SALT3.FRAG surfaces are compared in Figure~\ref{m0-m1} ($M_0$ and $M_1$) and Figure~\ref{cL} ($C_L$). The $M_0$ component at peak phase (middle panel of Figure~\ref{m0-m1}) is largely consistent between SALT2.FRAG and SALT3.FRAG, with slight differences in feature depths seen at $\lambda_{\rm rest} \lessapprox 4000$\AA. We note that the difference in $M_0$ components is more significant at the phases further from peak brightness.

\begin{figure*}
  \begin{center}
   \includegraphics[width=1.55\columnwidth]{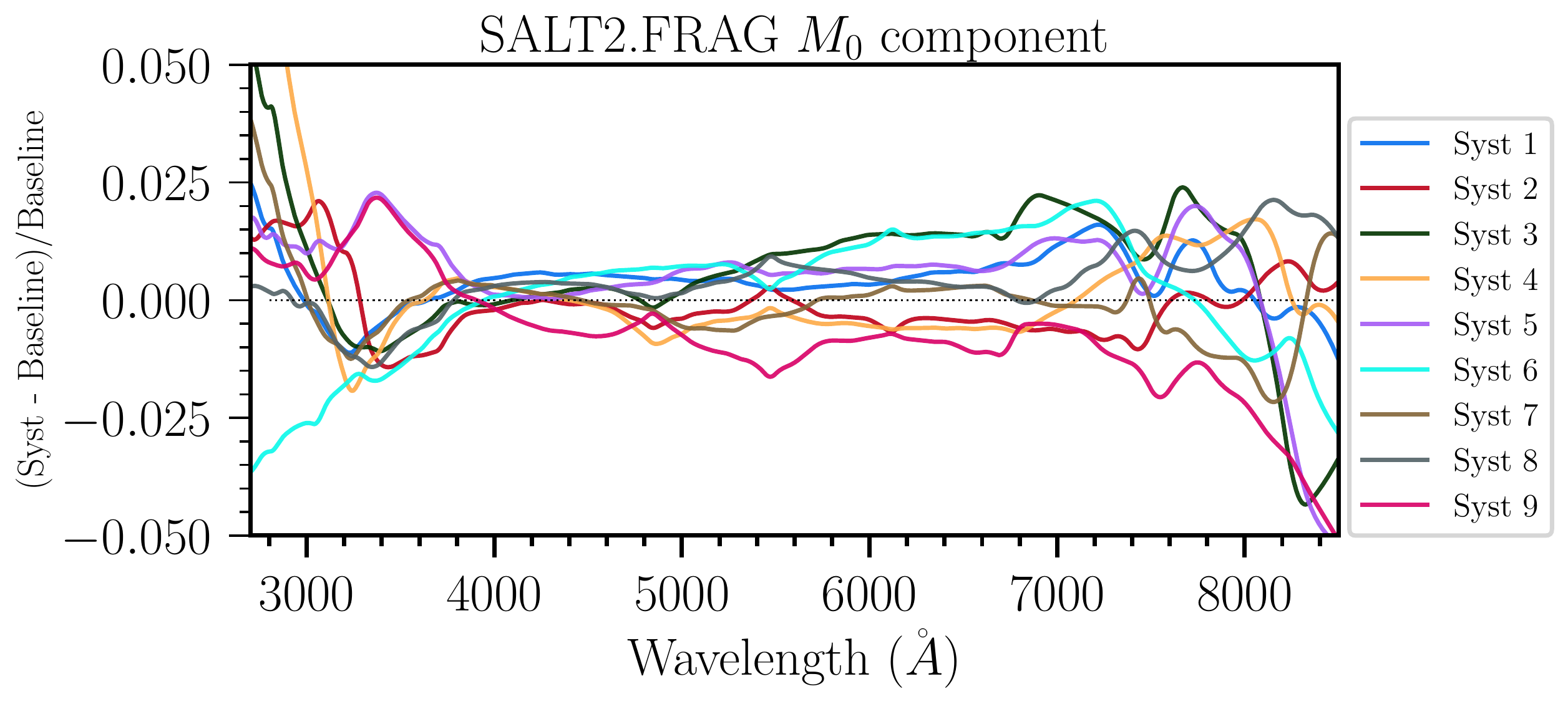}
   \includegraphics[width=1.55\columnwidth]{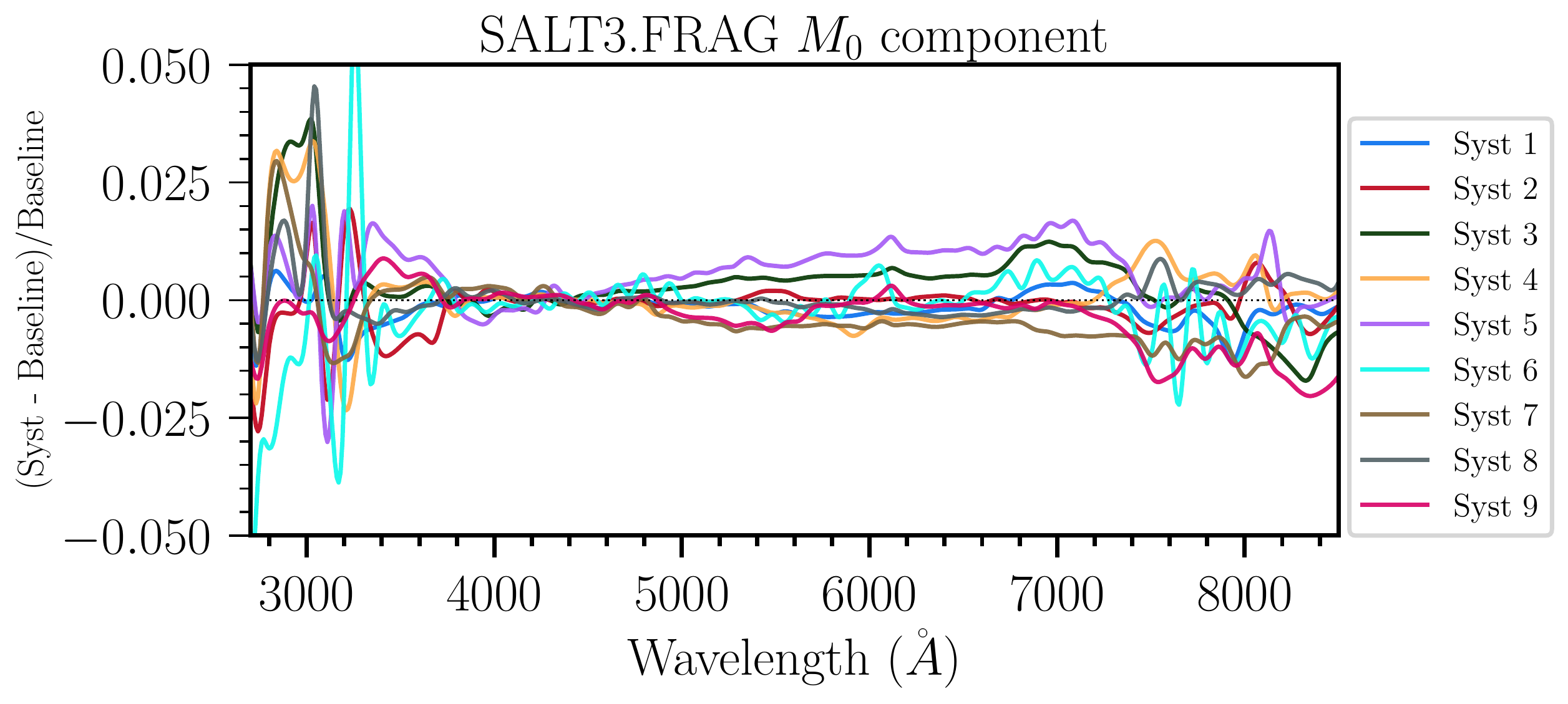}
  \end{center}
  \raggedleft
   \includegraphics[width=1.8\columnwidth]{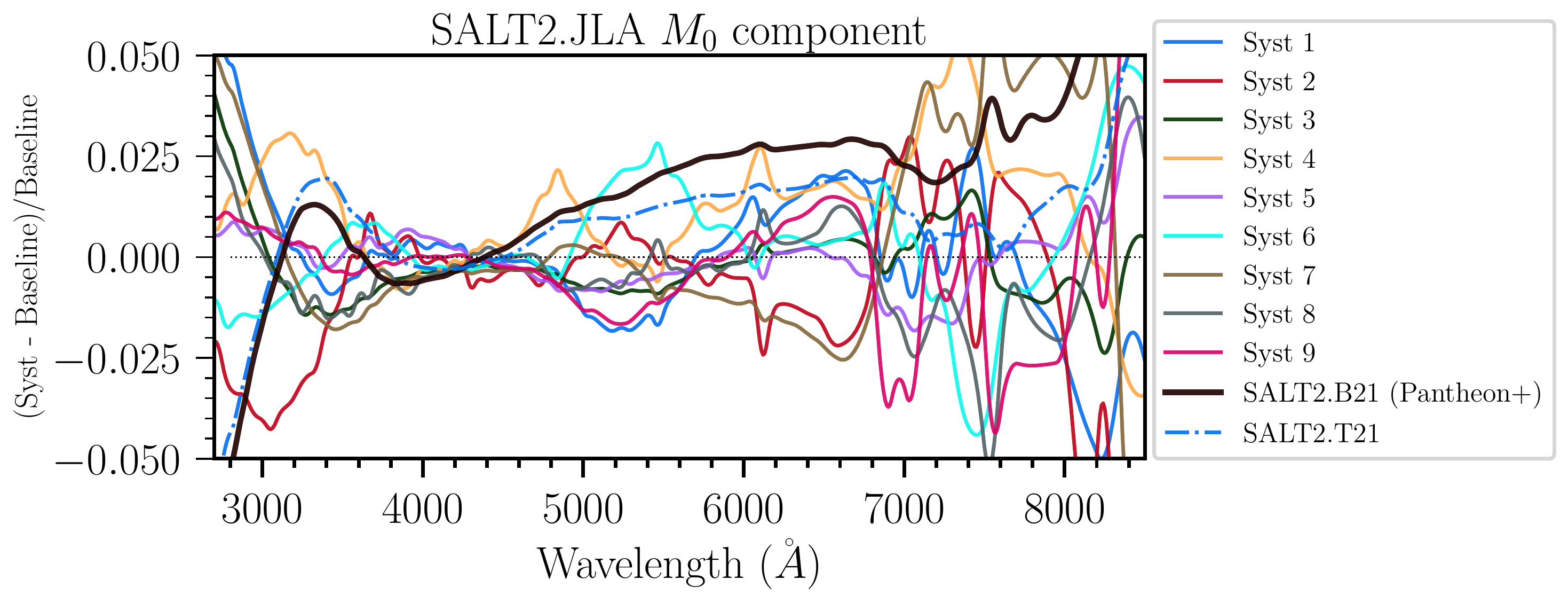}
  \caption{The impact of systematic uncertainty surfaces on the $M_0$ component of their respective base SALT surfaces, at maximum light. Each systematic surface is trained with a random set of zero point and central filter wavelength perturbations applied to the base surface inputs, drawn from the expected uncertainties in the photometric systems. Thus, these systematic surfaces probe the impact of photometric calibration errors on each base SALT surface.\\
  \textit{Top:} Systematic uncertainty surfaces of the base SALT2.FRAG surface (with FRAG perturbations applied). 
  \textit{Middle:} Systematic uncertainty surfaces of the base SALT3.FRAG surface (with FRAG perturbations applied). 
  \textit{Bottom:} Systematic uncertainty surfaces of the base \citealt{jla} (SALT2.JLA) surface (with JLA perturbations applied). For comparison, we also present the SALT2 M0 component from \citealt{T21} (SALT2.T21) and \citealt{superfrag} (SALT2.B21), which use the same training sample as SALT2.JLA but utilise updated calibration schemes.}
  \label{SALT23_sys}
\end{figure*}

Larger differences ($\pm 3\sigma$ from mean) are seen in the $M_1$ component at extreme values of stretch (upper and lower panels of Figure~\ref{m0-m1}). This is due in-part to SALT3's additional requirement that stretch is independent from colour, which affects the training of $M_1$. The effect of the different regularization schemes between the models can also be seen --- SALT3 consistently displays a flatter SED at $\lambda\lessapprox2800$\AA, and reduces the amount of unphysical negative flux. This difference is likely due to SALT3's more constraining regularization in the near-UV.

The SALT2.JLA surface from \citetalias{jla} is also shown, as it is the most commonly used surface from recent SN cosmology analyses. The biggest difference between SALT2.JLA and SALT2.FRAG/SALT3.FRAG is in the 3000-4000\AA\ wavelength range, where the FRAG surfaces are up to 5\% brighter than JLA. This effect, which is almost entirely covered by the model uncertainty, comes directly from the K21+ training sample, which is $\sim2.5\times$ larger than that used to train JLA.

The colour laws are largely consistent between SALT2.FRAG and SALT3.FRAG, despite the differences in $CL$ definition described in \S~\ref{section:23_differences}. The largest differences are seen at the endpoints, where the linear extrapolation scales up the small differences in the defined polynomials. The published SALT2.JLA colour law is consistent with SALT3.FRAG at low wavelengths and SALT2.FRAG at high wavelengths, though we refer to $\S$~\ref{section:SALT2-training} and Appendix~\ref{app:cl} for a discussion of issues in the JLA law.

\subsection{Systematic Uncertainty Surfaces}
\label{subsec:systsurf}

We test the sensitivity of the SALT2.FRAG and SALT3.FRAG surfaces to photometric calibration errors by training a suite of 9 "systematic uncertainty surfaces" for each model. Each systematic surface is trained with a randomly generated zero-point offset and central filter wavelength shift (drawn from the expected uncertainties in the photometric systems) applied to each of the filters used to observe the K21+ training sample. Measured correlations are included in the random shifts. The perturbations applied to each systematic uncertainty surface are identical for SALT2.FRAG and SALT3.FRAG, and are the same as the set used in \citet{superfrag}, except for those applied to the CSP $u$-band (here, we apply a correction to the previously used perturbations, which were found to be underestimated). The suites of systematic uncertainty surfaces are publicly available online.\footnote{\url{10.5281/zenodo.7400436}}

Surfaces that are more sensitive to photometric calibration uncertainties will experience a larger change in their trained components and/or uncertainties, which will propagate through to the fitted light curve parameters and distances. Changes in calibration have been shown to alter the $M_0$ component of the SALT models more than $M_1$ or $CL$, and changes in the $M_0$ component have been identified as the strongest driver of changes in $w$ compared to $M_1$ or $CL$ \citep{superfrag}. In Figure~\ref{SALT23_sys} we plot the fractional changes in $M_0$ versus wavelength for the SALT2.FRAG and SALT3.FRAG systematic suites. We also show the SALT2.JLA changes (\citealt{jla}) for comparison. The perturbations applied in the SALT2.JLA systematic suite are not exactly the same as those in the SALT2.FRAG and SALT3.FRAG suites --- the JLA perturbation values (which were not publicly released) had to cover a smaller set of photometric systems in the training sample, so in order to train new systematic uncertainty suites new perturbation values were defined (details are given in Appendix A of  \citealt{superfrag}). While we cannot directly compare each systematic FRAG surface to the corresponding JLA surface, we can learn about the change in photometric calibration uncertainty by assessing each suite in its entirety.

The level of variation in the panels of Figure~\ref{SALT23_sys} show that both of our SALT.FRAG surfaces are less sensitive to systematic errors in photometric calibration than SALT2.JLA, because they are constrained with more training sample data. The base SALT2.JLA $M_0$ surface changes by at least 5\% with some systematic uncertainty surfaces, for wavelength regions below $\approx3000$\AA\ and above $\approx7400$\AA. While SALT2.FRAG and SALT3.FRAG experience some $M_0$ shifts on order of 5\% for wavelengths below $\approx3000$\AA, the shift reduces to a maximum of $\approx2.5$\% for regions above $\approx3000$\AA. 
We note that across all wavelengths, SALT3.FRAG is less sensitive to photometric uncertainties than SALT2.FRAG. The appearance that SALT3.FRAG's $M_0$ variation goes to 0 at $\sim4400$\AA\ is not fundamental; there is in fact variation at all wavelengths, but it is not always visible in Figure~\ref{SALT23_sys} due to the scale of the plot.

\section{A Cosmology Analysis with SALT.FRAG Surfaces}
\label{section:analysis}

\subsection{Description of Cosmology Sample Data}

The cosmology sample we use to test the performance of our surfaces is the public sample from the Dark Energy Survey's 3-Year Supernova Analysis (DES-SN3YR, \citealt{Abbott_2019, des3yr_systematics}).\footnote{Available from \url{https://des.ncsa.illinois.edu/releases/sn}.} This sample includes 207 spectroscopically confirmed SNe~Ia from the first three years of DES (\citealt{des_diffimg, smith_20}),
combined with a selection of 122 low-$z$ SNe~Ia ($0.01~<~z~<~0.1$). The low-$z$ subset includes SNe~Ia from the Harvard-Smithsonian Center for Astrophysics surveys (CfA3, CfA4; \citealt{Hicken_2009_cfa3}, \citealt{cfa4}) and the Carnegie Supernova Project (CSP; \citealt{csp}, \citealt{csp_2}).

\subsection{Methodology}
\label{methods}

We perform the light curve fitting, simulated bias corrections, and cosmology fitting described here using \verb|SNANA| \citep{snana09, Kessler_2019_sims}, a publicly-available supernova analysis software package. We use the method of \citet{salt2mu} to determine the standardization parameters $\alpha$ and $\beta$, and \verb|BBC| \citep{Kessler_2017} for the one-dimensional (redshift-dependent) bias corrections. 
We use SNANA's fast cosmology fitting program (which uses the \verb|MINUIT| $\chi^2$ minimisation of \citealt{minuit}) to fit for $\Omega_m$ and $w$, applying a cosmic microwave background (CMB) prior (described further on). The analysis workflow is controlled by \verb|Pippin| \citep{Hinton2020} pipeline for SN analyses. 

\subsubsection*{Methods for Simulations}

We simulate realistic sets of DES-SN3YR-like SNe~Ia photometry for three purposes: bias correcting distance measurements (bias-sims), comparing fitted parameter values to true values (scatter-sims), and empirically measuring the scatter in our final $\Delta w$ measurement (scatter-sims). Following Figure~1 of \citet{Kessler_2019_sims}, our simulation process generates SN~Ia SEDs in the rest-frame, applies realistic astrophysical effects and redshifts to the SED to simulate the true SED at the top of Earth's atmosphere, integrates the flux over each filter bandpass, adds Poisson noise based on observation characteristics, and writes simulated events that would be detected and spectroscopically confirmed by DES and Low-$z$ surveys (based on selection functions). The bias-sims and scatter-sims are generated twice, once using SALT2.FRAG as the model for SED generation, and once using SALT3.FRAG. For each surface, we simulate $\sim$150,000 SNe~Ia for the bias-sims, and $50$ samples of $\sim$500 SNe~Ia each for the scatter-sims (which are trimmed to 329 SNe further along the analysis pipeline).

Our simulations are generated using 
$\Omega_m = 0.315$, $w = -1.00$, $H_0 = 70.0$ ($\Lambda_{\rm CDM}$). To simulate intrinsic SED brightness variations, we use the G10 model from \citet{Kessler_2013} that is based on the model dispersion in \citet{g10_scatter}. Our simulations do not include the in-sample variance of the SALT surfaces (discussed in $\S$~\ref{section:23_differences}) when modelling the SN intrinsic variability, though this variance \textit{is} used as part of the model uncertainty later in the light-curve fitting stage. The true stretch and colour distributions are determined using the methods of \citet{Popovic_2021}, which constrains simulated and data distributions to match after selection requirements. These parent populations are given in Appendix~\ref{app:parentpops}.

\subsubsection*{Methods for Light Curve Fitting}

We perform light curve fits to the DES-SN3YR data and simulations with both SALT2.FRAG and SALT3.FRAG, and compare the fitted SN parameters $x_0$, $x_1,$ and $c$ from both. The valid SED wavelength range over which SALT models are considered to be accurate varies depending on the training sample and framework. The wavelength range used for light curve fitting is based on the central wavelengths of the filters with which the data were observed; the rest-frame central wavelength $\lambda_{\rm cen}$ of a filter must fall within this range in order for that filter to be used in the light curve fit. The $\lambda_{\rm cen}$ range that we use must be narrower than the total SED range (given in $\S$~\ref{section:23_differences}), because the entire bandpass (transformed to rest-frame) must be contained by SED wave range.

To make a fair comparison between surfaces,
we fit light curves over a rest-frame central filter wavelength range of 3000 $\leq \lambda_{\rm cen} \leq$ 8000\AA. We select this range because:

1) The blue ends (2000-3000\AA) of the SALT2.FRAG and SALT3.FRAG models have limited training data (see Figure 10 of \citetalias{salt3}), so the models are less reliable in this rest-frame region. The 3000 $\leq \lambda_{\rm cen}$ requirement ensures that at least 50\% of the filter curve falls into the well-constrained region of the SALT SED. 

2) The additional K21+ training sample data allows us to extend the reliable central filter wavelength range of SALT2 from the previously used upper limit of $\lambda_{\rm cen} \leq$ 7000\AA. This extension enables us to fit DES $i$-band data at all redshifts (the previous limit restricted $i$-band data to $z\gtrsim0.1$). 
As the SALT2 SED is only defined to 9200\AA, it cannot make use of the full training power of the K21+ training sample, so we conservatively restrict the trusted upper limit to be about a filter-width ($\approx$1000\AA) from the model endpoint (i.e., $\lambda_{\rm cen} \leq$ 8000\AA). We verify that the SALT2 model covariances are reasonable over this wavelength range.

If we were not restricted by the need to compare with SALT2.FRAG we would use the SALT3.FRAG model to light curve fit over a redder range, extending to $\lambda_{\rm cen} \lesssim 8700$\AA\
to increase the redshift-range over which DES $z$-band data can be used.

\subsubsection*{Methods for Measuring Distances and Cosmological Parameters}

We use the light curve fits to measure distance moduli for DES-SN3YR according to Equation~\ref{mod_tripp}, applying one-dimensional (1D) bias corrections \citep{Kessler_2017}. 1D bias corrections measure the average $\mu$-bias ($\Bar{\delta}_{\mu}$) as a function of redshift by analysing large simulations with the SALT2mu distance fitting procedure, and then use the recovered biases to correct the distance moduli in the data for sample selection and light curve fitting effects ($\mu \xrightarrow[]{} \mu - \Bar{\delta}_\mu\left(z\right)$). 
We are using a spectroscopically-confirmed sample of SNe~Ia, hence P(Ia)=1.0 (i.e., the likelihood of an event in our sample being a core-collapse SN is zero), so we only employ the bias correction procedure from BBC and not the BEAMS procedure (which accounts for core-collapse contamination).

We compare the bias corrected distance moduli between surfaces, for both the simulations and the real data. During this stage, the scatter-sim samples are trimmed so that the number of SN for each sub-survey matches those in the public DES-SN3YR sample. 

\begin{figure*}
  \begin{center}
   \includegraphics[width=1.6\columnwidth]{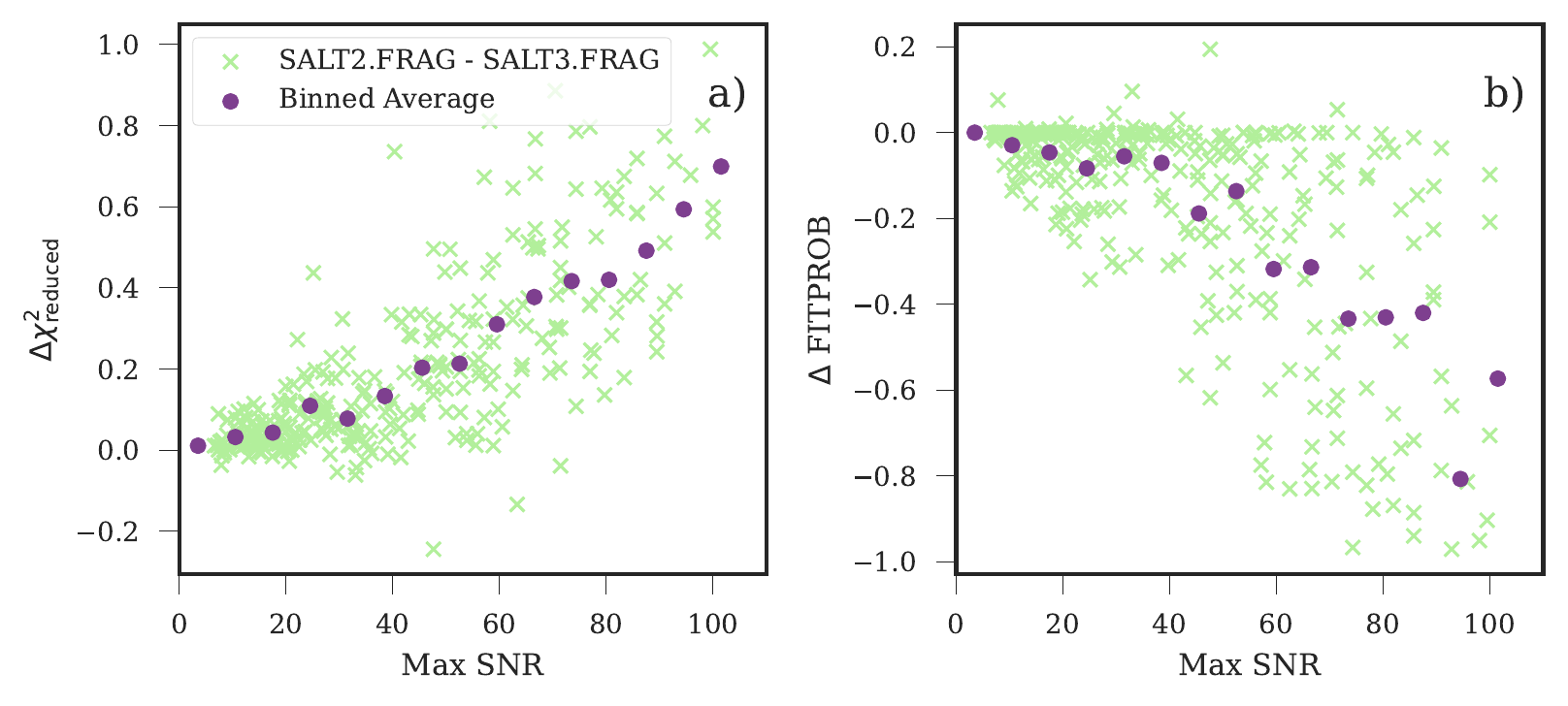}
  \end{center}
  \caption{Correlations between maximum signal-to-noise (SNR), reduced $\chi^2$, and fit probability when fitting DES-SN3YR with SALT2.FRAG and SALT3.FRAG. We plot the differences between the SALT2.FRAG and SALT3.FRAG fits to the DES-SN3YR data.}
  \label{data_deltafitprob}
\end{figure*}

\begin{table*}
\renewcommand\thempfootnote{\arabic{mpfootnote}}
\begin{minipage}{0.9\textwidth}
  \renewcommand{\arraystretch}{1.5}
  \centering
  \caption{Number of DES-SN3YR SNe~Ia that failed light curve fitting cuts for each SALT*.FRAG surface, for a variety of criteria. SNe that failed cuts with one surface but passed cuts with the other surface are shown in parentheses. The right-most column shows the number of overlapping SN remaining after cuts for both surfaces are applied. \\
  \textit{Trestmin = rest-frame phase of earliest observation epoch; Trestmax = rest-frame phase of latest observation epoch.}}
    \label{nsn_table}
    \begin{tabular}{|l|l|c|c|c|}
    \hline
     &  & \textbf{\begin{tabular}[c]{@{}c@{}}Cut with SALT2.FRAG \\ (but passed with SALT3.FRAG)\end{tabular}} & \textbf{\begin{tabular}[c]{@{}c@{}}Cut with SALT3.FRAG \\ (but passed with SALT2.FRAG)\end{tabular}} & \textbf{Common SN remaining} \\ \hline
    \multirow{4}{*}{\textbf{Low-Z}} & \textit{$-0.3 < c < 0.3$} & 0 & 1 & \multirow{4}{*}{105} \\ 
     & \textit{$-3.0 < x_1 < 3.0$} & 1 & 1 &  \\ 
     & \textit{$-20 < $ Trestmin $ < 0$} & 2 & 2 &  \\ 
     & \textit{SNR $> 5$ in $2$ bands} & 1 & 1 &  \\ 
     & \textit{Fit probability $> 0.01$} & 13 (12) & 0 &  \\ \hline
    \multirow{3}{*}{\textbf{DES}} & \textit{$-0.3 < c < 0.3 $} & 1 & 1 & \multirow{3}{*}{203} \\ 
     & \textit{$10 <$ Trestmax $< 200$} & 0 & 1 (1) &  \\ 
     & \textit{Fit probability $> 0.01$} & 2 (2) & 0 &  \\ \hline
\end{tabular}
\end{minipage}
\end{table*}

Finally, we estimate the impact of the choice of SALT2.FRAG or SALT3.FRAG on cosmological parameters by fitting $\Omega_m$ and $w$, assuming a wCDM model, i.e. a flat universe with a constant $w$ value and cold dark matter. 
We apply an approximate CMB prior with $\rm R_{CMB} = 1.757$,\footnote{R-shift parameter described in e.g., Eq. 69 of \citet{wmap_5}} $\sigma_{\rm R_{CMB}} = 0.007$, tuned such that the constraining power is similar to that of Planck \citep{planck2015}. We also fit "SN-only" values of $w$ by removing the Planck prior and applying an uninformative flat prior for $\Omega_m$, to evaluate the SALT framework's impact on the SN data alone.


We calculate $\Delta w = w_{\rm SALT2.FRAG} - w_{\rm SALT3.FRAG}$. We estimate the uncertainty on $\Delta w$ for the simulations by running the cosmology analysis on each of the 50 scatter sims and calculating the standard error $\sigma_{\Delta w}$ on the mean of those $\Delta w$ values. We assume that the simulated $\sigma_{\Delta w}$ is also valid for the real data.

\subsection{Results - Real DES-SN3YR Data}
\label{section:real_results}

\subsubsection*{SNe~Ia That Pass Light Curve Fitting Cuts}

As we are using the public DES-SN3YR sample, their cuts have already been applied.\footnote{See Table~1 of \citet{des3yr_systematics}.} This means that our cuts can remove events, but we cannot recover events from the DES-SN3YR sample. We note that 276 of the DES-SN3YR SNe are present in the K21+ sample, so our surfaces have been trained on data that they are now fitting. 

Of the original 329 DES-SN3YR events, 309 
SNe pass light curve fitting cuts with SALT2.FRAG, and 322 
SNe pass cuts with SALT3.FRAG. A total of 308 common SNe pass cuts using both SALT2.FRAG and SALT3.FRAG.
The selection requirements (cuts) are summarised in Table~\ref{nsn_table}.

The majority of SNe are cut due to the fit probability > 0.01 criterion when fitting the low-$z$ sample with SALT2.FRAG. The fit probability is the probability of finding an equal or larger light curve data-model $\chi^2$, assuming Gaussian-distributed flux uncertainties. The low-$z$ sample contains the majority of the high signal-to-noise (SNR) SNe in the DES-SN3YR sample. 
In Figure~\ref{data_deltafitprob}a, we show that the change between the reduced $\chi^2$ (i.e. $\chi^2 / \rm NDOF$) of SALT2.FRAG
and SALT3.FRAG is strongly correlated with maximum SNR\footnote{Maximum SNR is defined as the maximum signal-to-noise ratio among all observations used in the light curve fit of a single SN.}, as is the change in fit probability in Figure~\ref{data_deltafitprob}b. This effect can be attributed to the model uncertainties --- SALT3 appears to have larger model uncertainties, which lowers the reduced $\chi^2$ of high-SNR data. We examine this effect further in this work using simulated data.

The single SN that is cut with SALT3.FRAG but selected with SALT2.FRAG has an absolute Hubble residual\footnote{Hubble residual distance modulus inferred from Equation~\ref{mod_tripp} minus distance modulus at the same redshift from the best-fit $w$CDM (using all SNe, without CMB priors.)} of $2.6\sigma$. The 14 SNe that are cut with SALT2.FRAG but selected with SALT3.FRAG have a mean absolute Hubble residual of $0.66\sigma$.
The Hubble scatter is $\sigma = 0.152$ mag for SALT2.FRAG and $\sigma = 0.151$ mag for SALT3.FRAG.

\subsubsection*{Light Curve Parameters and Distance Measures}

\begin{figure*}
  \begin{center}
   \includegraphics[width=1.8\columnwidth]{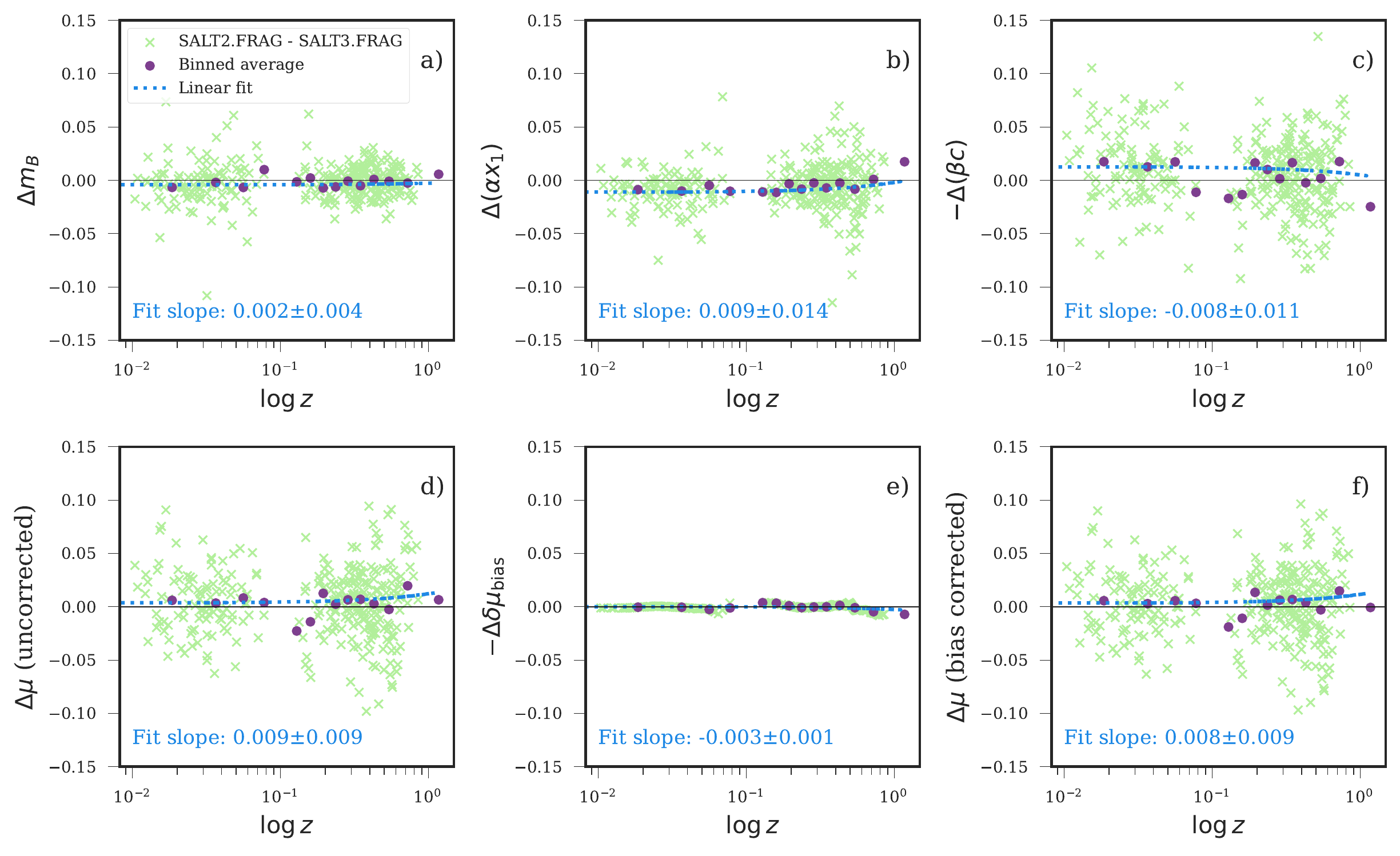}
  \end{center}
  \caption[]{Differences in fitted parameters (for DES-SN3YR data) between the SALT2.FRAG and SALT3.FRAG surfaces, plotted versus log redshift. Differences are given as e.g., $\Delta \mu = \mu_{SALT2} - \mu_{SALT3}$. The nuisance values $\alpha$ and $\beta$, which are unique to each surface, are fitted parameters from the \texttt{BBC} process when each surface was allowed to use its full set of fitted data, though only the common SNe are plotted here. Each SN is plotted as a green cross, with the binned averages plotted as purple circles. The bins match those used by \texttt{BBC}. The linear lines of best-fit $y = mx + c$ are shown as blue dotted lines, with the fit slope $m$ reported in each panel.\\
  Each term is plotted with its sign according to the Tripp equation, so that the true impact on $\mu$ is shown in each panel. The top row shows the (uncorrected) supernova parameters: $m_B$, $\alpha~\times$~stretch, and $- \beta~\times$~colour. 
  Together with $M$ (where $\Delta M_{\rm avg} = -0.006$), these sum to the (pre-bias corrected) distance modulus.
  The bottom row shows: pre-bias corrected $\Delta \mu$, the change in 1-D bias corrections applied to $\mu$, and $\Delta \mu$ after bias corrections.
  } 
  \label{sixpanel_data}
\end{figure*}

\begin{table*}
\renewcommand\thempfootnote{\arabic{mpfootnote}}
\begin{minipage}{0.9\textwidth}
\renewcommand{\arraystretch}{1.5}
\centering
\caption{Cosmological parameters from \texttt{wfit} for various runs of the analysis, with SALT2.FRAG and SALT3.FRAG. The "Cosmology measurement" column refers to the information used in the cosmology fit; that is, the SN data with either no priors (\textit{SN}) or CMB-like priors (\textit{SN + CMB}).
For simulations we report the mean $\Delta w$ and standard error on the mean ($\sigma$) from 50 sets of simulations, as well as the dispersion (\textit{RMS, root-mean-square = $\sigma \times \sqrt{50}$}). The "$\sigma$ from real data" column shows how far the mean simulated data result is from the corresponding real data result, in units of the simulated standard error. The "$\sigma$ from 0" column shows how far the result is from $\Delta w = 0$.}
\label{deltaw_table}
\begin{tabular}{|c|l|l|l|c|c|c|c|}
\hline
\textbf{\begin{tabular}[c]{@{}l@{}}Cosmology\\ measurement\end{tabular}} 
& \textbf{\begin{tabular}[l]{@{}l@{}}Subsample used\\ for cosmology\end{tabular}} 
& \textbf{\begin{tabular}[l]{@{}l@{}}Sample used \\ for light curve fits\end{tabular}} 
& \textbf{$\Delta w$\footnote{$\Delta w = w_{\rm SALT2.FRAG} - w_{\rm SALT3.FRAG}$}} 
& $\sigma$
& RMS
& \textbf{$\sigma$ from real data}
& \textbf{$\sigma$ from 0} \\ \hline
\multirow{7}{*}{SN} & \multirow{4}{*}{All SNe} & DES-SN3YR data\footnote{SALT2.FRAG = 309 SNe, SALT3.FRAG = 322 SNe; see $\S$~\ref{section:real_results}.} & -0.032 & - & - & - & - \\ \cline{3-8} 
 &  & \cellcolor{pink!65} SALT2.FRAG-generated sims & +0.033 & 0.026 & 0.184 & $2.5\sigma$ & $1.3\sigma$ \\ \cline{3-8} 
 &  & \cellcolor{blue!15} SALT3.FRAG-generated sims & +0.008 & 0.035 & 0.247 & $1.1\sigma$ &  $0.2\sigma$ \\ \cline{3-8} 
 &  & \cellcolor{green!15} Independently-generated sims & -0.025 & 0.039 & 0.276 & - & $0.6\sigma$ \\ \cline{2-8} 
 & \multirow{3}{*}{Common SNe} & DES-SN3YR data & -0.010 & - & - & - & - \\ \cline{3-8} 
 &  & \cellcolor{pink!65} SALT2.FRAG-generated sims & -0.007 & 0.007 & 0.049 & $0.4\sigma$ &  $1.0\sigma$ \\ \cline{3-8} 
 &  & \cellcolor{blue!15} SALT3.FRAG-generated sims & -0.028 & 0.006 & 0.042 & $3.0\sigma$ &  $4.6\sigma$ \\ \hline
\multirow{7}{*}{SN + CMB} & \multirow{4}{*}{All SNe} & DES-SN3YR data & +0.001 & - & - & - & - \\ \cline{3-8} 
 &  & \cellcolor{pink!65} SALT2.FRAG-generated sims & -0.004 & 0.005 & 0.035 & $1.0\sigma$ &  $0.8\sigma$ \\ \cline{3-8} 
 &  & \cellcolor{blue!15} SALT3.FRAG-generated sims & -0.005 & 0.004 & 0.028 & $1.5\sigma$ &  $1.3\sigma$ \\ \cline{3-8} 
 &  & \cellcolor{green!15} Independently-generated sims & +0.005 & 0.007 & 0.049 & - &  $0.7\sigma$ \\ \cline{2-8} 
 & \multirow{3}{*}{Common SNe} & DES-SN3YR data & +0.008 & - & - & - & -\\ \cline{3-8} 
 &  & \cellcolor{pink!65} SALT2.FRAG-generated sims & -0.011 & 0.001 & 0.007 & $19\sigma$ &  $11\sigma$ \\ \cline{3-8} 
 &  & \cellcolor{blue!15} SALT3.FRAG-generated sims & -0.012 & 0.001 & 0.007 & $20\sigma$ &  $12\sigma$\\ \hline
\end{tabular}
\end{minipage}
\end{table*}

We fit light curves to the DES-SN3YR data using each surface, and compare the resultant differences in fitted light curve parameters in Figure~\ref{sixpanel_data}a - \ref{sixpanel_data}c. We only plot the 308 SN that are common to both surfaces. The binned averages show that the surfaces agree in each redshift bin used for cosmology to within at least 0.025 magnitudes for $m_B$, $\alpha x_1$, and $\beta c$. We note that on average, SALT3.FRAG fits a slightly higher $\alpha x_1$ value than SALT2.FRAG (resulting in a negative residual in Figure~\ref{sixpanel_data}b); we confirm that this difference originates from the fitted $x_1$ values rather than the $\alpha$ values.\footnote{$\Delta \alpha$ = $7 \times 10^{-4}$ (unitless)} The impact of these higher SALT3 $\alpha x_1$ values is somewhat cancelled out by SALT3 also fitting higher (on average) $\beta c$ values (resulting in the positive residual shown in Figure~\ref{sixpanel_data}c, as we plot $-\beta c$ according to the Tripp equation); we confirm that this difference once again originates from the fitted $c$ values rather than the $\beta$ values.\footnote{$\Delta \beta$ = $0.12$ (unitless)}

We compute and apply 1D-bias corrections for distance moduli using \verb|BBC|. The distance modulus difference before bias corrections is shown in Figure~\ref{sixpanel_data}d, the bias correction difference is shown in Figure~\ref{sixpanel_data}e, and the final difference in bias-corrected distance is shown in Figure~\ref{sixpanel_data}f. As follows from the light curve parameter results, all DES-SN3YR SNe bias-corrected distance moduli fit with SALT2.FRAG and SALT3.FRAG agree to within $\approx~0.1$~mag - and the average difference per bin is at least an order of magnitude smaller. The slope of the change in the bias-corrected distance modulus with respect to redshift is consistent with 0 to $<1\sigma$, indicating that the choice of SALT model framework does not bias the DES-SN3YR cosmology.

We inspect the distribution of fitted parameter uncertainties for the two surfaces, using the common sample of 308 SN. These results are similar to those obtained using the full sample for each surface. The mean $\sigma_{m_B}$ is comparable between SALT2.FRAG and SALT3.FRAG, with a mean value of 0.044~mag and 0.045~mag respectively. The mean $\sigma_{x_1}$ is smaller with SALT2.FRAG (0.333~mag versus 0.378~mag), as is the mean $\sigma_{\alpha x_1}$ (0.052~mag versus 0.059~mag). However, the mean $\sigma_c$ is larger with SALT2.FRAG (0.035~mag versus 0.032~mag), as is the mean $\sigma_{\beta c}$ (0.099~mag versus 0.093~mag). This propagates through to a mean $\sigma_\mu$ that is smaller with SALT2.FRAG (0.153~mag versus 0.156~mag).

\subsubsection*{Cosmology}

Fitted cosmological parameters for different sub-samples of DES-SN3YR are shown in Table~\ref{deltaw_table}, for both SN-only and SM+CMB fits. When both SALT surfaces are applied to their full sample of SNe~Ia that pass cuts (i.e., 309 SNe from SALT2.FRAG, 322 SNe from SALT3.FRAG), we recover $\Delta w = -0.032$ using the SN data alone, and $\Delta w = +0.001$ when including a CMB-like prior.

The statistical uncertainty on the fitted $w$ value ($\sigma_w$) for either surface is an order of magnitude greater than the $\Delta w$ value ($\sigma_{w{(SN)}} \sim 0.2$; $\sigma_{w{(SN+CMB)}} \sim 0.05$), indicating that the choice of surface will not alter the DES-SN3YR cosmology results. As the uncertainty on each measurement is not independent, we need realistic simulations to estimate the uncertainty on $\Delta w$ ($\sigma_{\Delta w}$). This is done in $\S$~\ref{sim_results}. 

We also investigate the impact on cosmology of the different SN samples that passed cuts with each surface. We perform this test by removing any SNe that are not in the intersection of the SALT2.FRAG and SALT3.FRAG samples to leave a consistent fitted data sample of 308 SNe~Ia, re-computing the BBC nuisance parameters and distance measurements, and re-fitting the cosmology ("Common SNe" rows of Table~\ref{deltaw_table}). These results are discussed in more detail in $\S$~\ref{sim_results} and $\S$~\ref{sec:discussion}, after we attempt to use simulations to estimate their statistical significance.

\begin{figure*}
  \begin{center}
  \includegraphics[width=1.3\columnwidth]{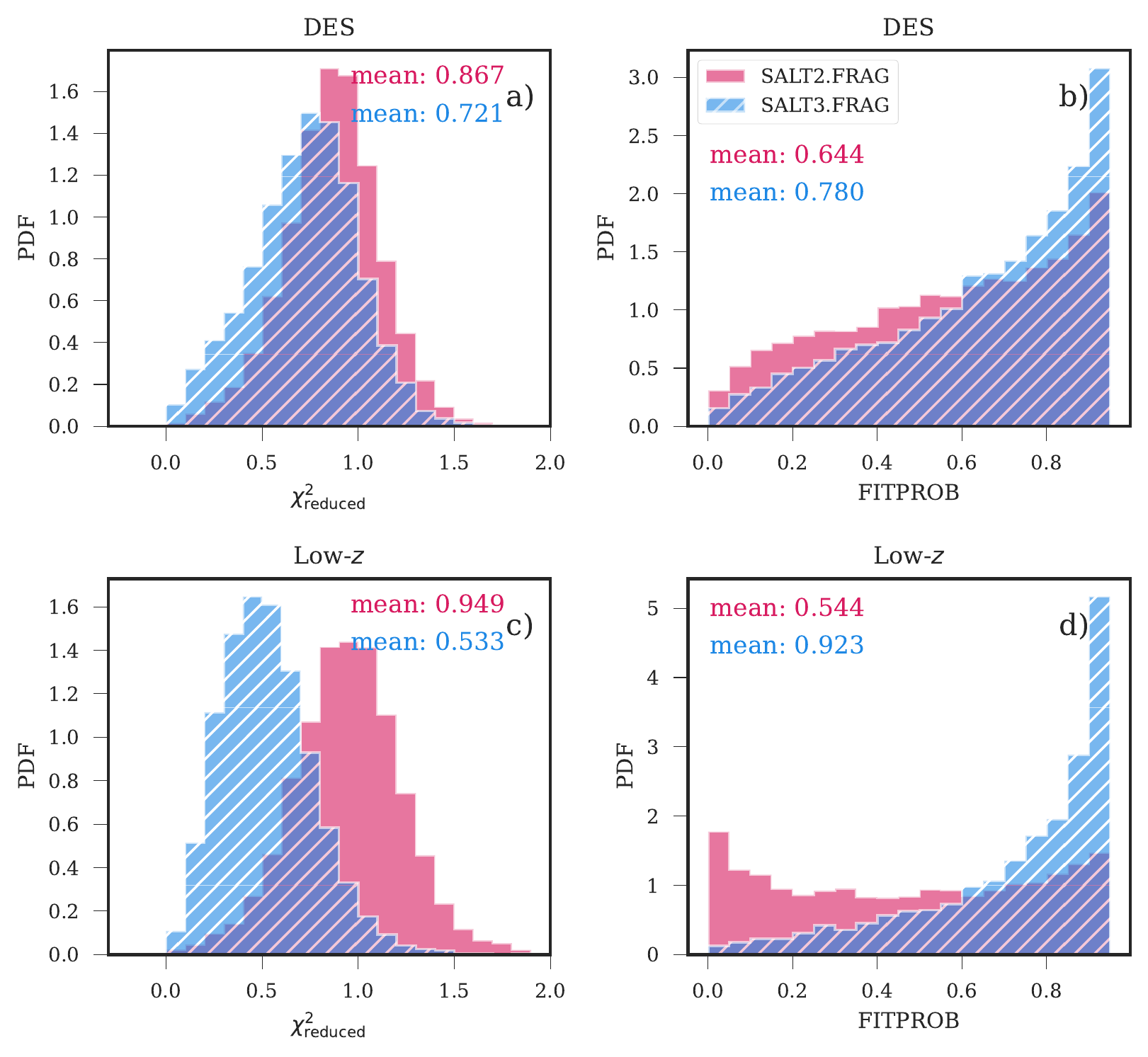}
  \end{center}
  \caption{Reduced $\chi^2$ and fit probability distributions for the full DES (\textit{top row}) and low-$z$ (\textit{bottom row}) fitted samples of simulated SNe~Ia for SALT2.FRAG and SALT3.FRAG. Here, the SALT2.FRAG results are simulated and fit with the SALT2.FRAG surface, and the SALT3.FRAG results are simulated and fit with the SALT3.FRAG surface, so each set of simulations is independent. These distributions agree with those obtained from fitting the real DES-SN3YR data, and with those obtained from fitting either surface to SALT2.FRAG- or SALT3.FRAG-generated simulations (for both the full and consistently fit samples).}
  \label{fitprob_simdist}
\end{figure*}

\subsection{Results - Simulated DES-SN3YR data}
\label{sim_results}

In this section, we report three sets of simulated results:

\begin{itemize}
    \item Independent simulated samples, where SALT2.FRAG is used to generate and fit a simulated sample, and SALT3.FRAG is used to generate and fit a separate simulated sample. The underlying simulated samples here are independent for each surface.
    \item SALT2.FRAG generated simulations, where SALT2.FRAG is used to generate a simulation that is fit with both SALT2.FRAG and SALT3.FRAG. The underlying simulated sample is the same for both surfaces.    
    \item SALT3.FRAG generated simulations, where SALT3.FRAG is used to generate a simulation that is fit with both SALT2.FRAG and SALT3.FRAG. The underlying simulated sample is the same for both surfaces.   
\end{itemize}

For each, we simulate 50 $\times$ DES-SN3YR-like samples, and analyze each sample the same way as for real data. When examining the changes in fitted light curve parameters and distances, we group all 50 simulations together, and plot the results of the independent simulated samples to see the combined effects of simulating and light curve fitting with different surfaces (noting where the results from the other sets differ). The increased statistics of these combined samples allow us to study the results (e.g.,distributions of fitted parameters) in greater detail. When examining the changes in cosmology, we calculate $w$ for each of the 50 simulations individually, and report the mean of the $\Delta w$ for each set. We report $\sigma_{\Delta w}$ as the standard error on the mean. 

\subsubsection*{SNe~Ia That Pass Light Curve Fitting Cuts}

Regardless of the model used to generate simulations, SALT3.FRAG fits (on average) 1 more DES SN and 27 more low-$z$ SNe than SALT2.FRAG, for each of the 50 simulations. 

We find that the SNe that are cut with SALT3.FRAG have, on average, slightly higher Hubble residuals than those that are cut with SALT2.FRAG. This result agrees with the result obtained from the real data.

For the combined SALT2.FRAG generated simulations, the 126 SNe that are cut with SALT3.FRAG but selected with SALT2.FRAG have a mean absolute Hubble residual of $1.04 \sigma$.
The 1529 SNe that are cut with SALT2.FRAG but selected with SALT3.FRAG have a mean absolute Hubble residual of $0.83 \sigma$.
The Hubble scatter is $\sigma = 0.162$ mag for SALT2.FRAG and $\sigma = 0.163$ mag for SALT3.FRAG.

For the combined SALT3.FRAG generated simulations, the 110 SNe that are cut with SALT3.FRAG but selected with SALT2.FRAG have a mean absolute Hubble residual of $1.145 \sigma$.

The 1560 SNe that are cut with SALT2.FRAG but selected with SALT3.FRAG have a mean absolute Hubble residual of $0.83\sigma$.
The Hubble scatter is $\sigma = 0.160$ mag for SALT2.FRAG and $\sigma = 0.158$ mag for SALT3.FRAG.


While the statistics for DES SNe are comparable, SALT3.FRAG recovers a significantly higher number of low-$z$ SNe; a sample which provides a crucial anchor for cosmology. As seen in the real data, the majority of these low-$z$ cuts occur due to the fit probability > 0.01 criterion, and the change in fit probability is correlated with maximum SNR. Some of this observed difference in low-$z$ fit probability could feasibly arise from the difference in regularization schemes --- in the low-$z$ region, the CSP $u$- and CfA $U$-bands probe the near-UV rest frame wavelengths, where SALT3's regularization exhibits the largest differences compared to SALT2 (Figure~\ref{m0-m1}). However, this does not fully explain the effect.

We  plot the distributions of the reduced $\chi^2$ and fit probability in Figure~\ref{fitprob_simdist}, for the independent simulations. These distributions agree with those of the real data and the other sets of simulations.
When the reduced $\chi^2$ (which is used in the calculation of fit probability) is less than one, a model has "overfit" the data. For example, if the combined model-plus-data uncertainties are overestimated then the model will be too flexible, resulting in an artificially low $\chi^2$ (a discussion about the usefulness and appropriateness of reduced $\chi^2$ values for different astronomical models is given in \citealt{andrae_10}). Similarly, we expect the fit probability distribution to be approximately flat when the model-plus-data uncertainties are well understood, and to cluster around one for  overestimated uncertainties. 

We see both of these indicators of overestimated model-plus-data uncertainties in the distributions of Figure~\ref{fitprob_simdist}. Both SALT surfaces have fit probability distributions that peak around one for the DES sample. The SALT2.FRAG low-$z$ fit probability distribution appears somewhat flat, but the SALT3.FRAG distribution strongly peaks at one. As the photometry uncertainties in the low-$z$ sample are lower (as it has a higher average SNR), the increased model uncertainties from SALT3.FRAG have more impact in this sample than in DES (as follows from Figure~\ref{data_deltafitprob}).
The dependence of SALT2's fit probability \textit{distribution} on the data sample, seemingly correlated with SNR, has been an area of concern for many years. While the SALT3.FRAG model uncertainties may require fine-tuning, they do appear to resolve this issue of sample-dependent fit probability.

A preliminary comparison between SALT3.FRAG and a SALT3.FRAG-like surface with reduced model uncertainties gives a change in fit probability that trends with maximum SNR, indicating that at least a portion of low-$z$ sample recovery effect arises from the SALT3 model uncertainties. Including the in-sample variance in the simulations in order to fully model the intrinsic variability of SNe would also increase the reduced $\chi^2$ in Figure~\ref{fitprob_simdist}, although given we see similar reduced $\chi^2$ distributions in the real data, this also cannot fully explain the over-fitting.
While updating the simulation procedure is necessary to disentangle the effects of intrinsic scatter modelling versus model uncertainties, it is beyond the scope of this paper.

We examine the simulated (i.e., true) parameter distributions for the combined simulated samples fitted by each surface, remembering that as each SALT surface fits its own sample of simulated SNe~Ia, the underlying populations are not expected to be identical. The aforementioned low-$z$ sample recovery leads to SALT3.FRAG distributions with more low-$z$, low $m_B$ and low $\mu$ SNe compared to SALT2.FRAG. We also find a difference in $x_1$ distributions --- the SALT3.FRAG sample has a more prominent secondary peak at low stretch, which we find comes mainly from the underlying difference in parent populations in the simulations (not from the recovered populations after light curve fitting).

\begin{figure*}
  \begin{center}
   \includegraphics[width=2\columnwidth]{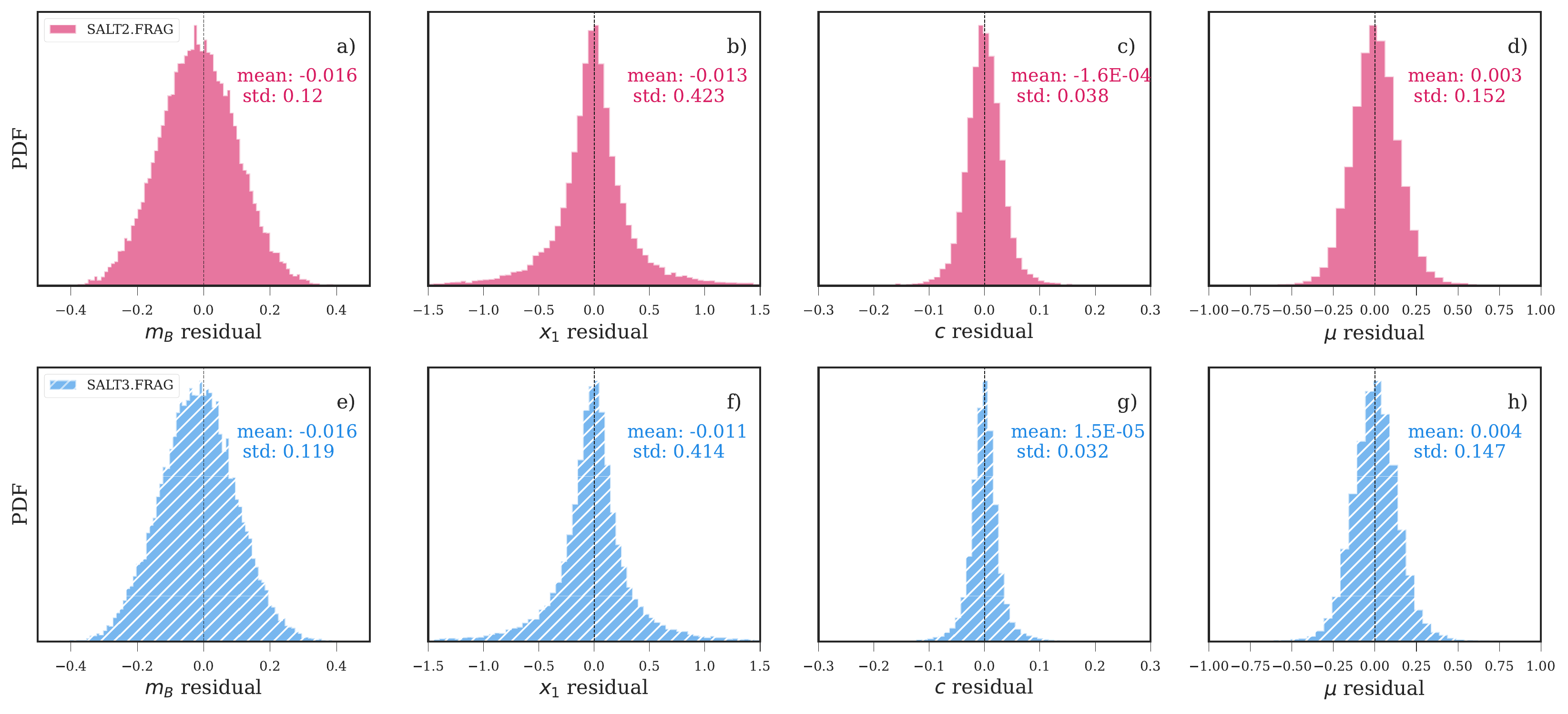}
  \end{center}
  \caption{Residual distributions of SN parameters for the full sample of simulated SNe~Ia fit with each surface. Here, the SALT2.FRAG results are from SNe that have been generated and fit with SALT2.FRAG (and similarly for SALT3.FRAG).}
  \label{sim_res_dist_sixpanel}
\end{figure*}

\begin{figure*}
  \begin{center}
   \includegraphics[width=2\columnwidth]{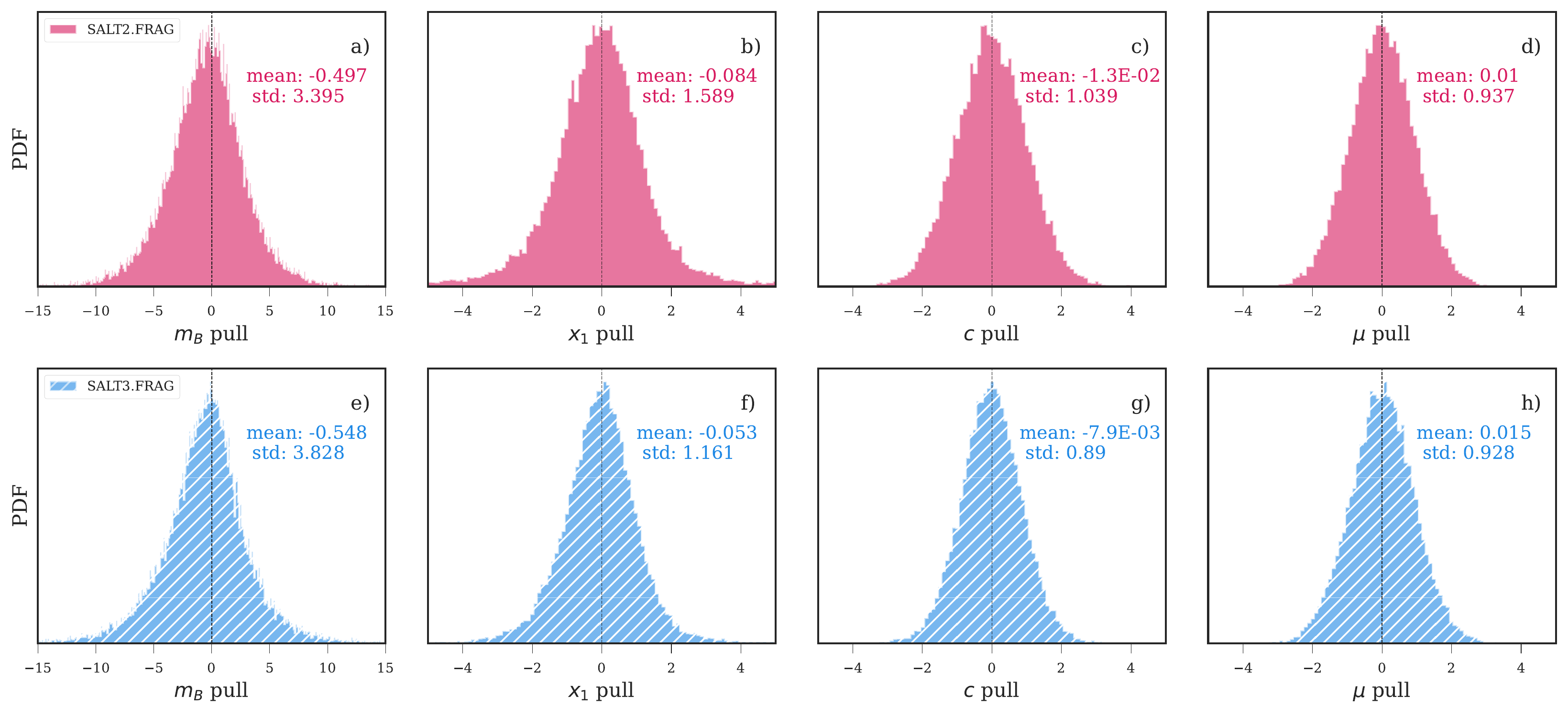}
  \end{center}
  \caption{Pull distributions of SN parameters for the full sample of simulated SNe~Ia fit with each surface.}
  \label{sim_pull_dist_sixpanel}
\end{figure*}

\subsubsection*{Light Curve Parameters and Distance Measures}

As each SALT surface fits its own independent sample of simulated SNe~Ia, we cannot compare the fitted parameters for each SN as we did for the real data. Instead, we use the simulations to probe how accurately each surface can recover the true light curve parameters. 

Figure~\ref{sim_res_dist_sixpanel} shows the distribution of residual values for each surface. Figure~\ref{sim_pull_dist_sixpanel} shows the distribution of pull values for each surface, defined as pull = (fit minus true)/fit error \citep{pulls}. The $m_B$ pull distribution is not centred around 0 for either surface; it is more non-zero and has a wider distribution with SALT3.FRAG light curve fits, regardless of the surface used to generate the simulations. The $x_1$ pull distribution is centred around 0 for either surface when fitting samples generated with the matching surface (as plotted), but becomes off-centre (mean $x_1 \sim 0.1$ mag) when fitting with simulations generated with the alternate surface. In all cases, the $x_1$ pull distribution is narrower with SALT3.FRAG. The $c$ pull distribution is centred around 0 for either surface, regardless of the surface used to generate the simulations. It is wider with SALT3.FRAG when fitting SALT2.FRAG-generated simulations, and narrower in all other cases. The $\mu$ pull distribution is centred around 0 for either surface and is consistently narrower with SALT3.FRAG, regardless of the surface used to generate the simulations.



We evaluate the precision of the SALT surfaces by inspecting the distribution of fitted errors. Our results echo what was seen in the real data sample, and unless otherwise stated, agree with the trends seen in the separate SALT2.FRAG- and SALT3.FRAG-generated simulations.

The $\sigma_{m_B}$ distribution is comparable between SALT2.FRAG and SALT3.FRAG, with a mean value of 0.043. The mean $\sigma_{x_1}$ is smaller with SALT2.FRAG (0.313 versus 0.348), as is the mean $\sigma_{\alpha x_1}$ (0.044 versus 0.048). However, the mean $\sigma_c$ is \textit{larger} with SALT2.FRAG (0.034 versus 0.031), as is the mean $\sigma_{\beta c}$ (0.108 versus 0.095). This propagates through to a mean $\sigma_\mu$ that is smaller with SALT2.FRAG (0.153 versus 0.156). The mean $\sigma_\mu$ distributions do change slightly when looking at results from the separate SALT2.FRAG- and SALT3.FRAG-generated simulations --- the mean $\sigma_\mu$ increases by 0.001 with SALT3.FRAG for simulations generated with SALT2.FRAG, and is equal for simulations generated with SALT3.FRAG.

\subsubsection*{Cosmology}

Cosmological parameter results from \verb|wfit| are shown in Table~\ref{deltaw_table}. To calculate these parameters, we first trimmed the simulations so that the number of SN for each sub-survey could not exceed the respective number of SN in the public DES-SN3YR sample. This allows us to compare our simulated cosmology results with those from the real data with minimal differences in statistical constraining power. 

We begin by looking at the results obtained when each surface is allowed to use all of the SNe that pass its light curve fitting cuts, i.e., the "All SNe" rows of Table~\ref{deltaw_table}. For the independent simulations we find a mean $\Delta w = -0.025 \pm 0.039$ when using SN data alone, and $\Delta w = 0.005 \pm 0.007$ when including a CMB-like prior.
When fitting the same underlying simulation with SALT2.FRAG and SALT3.FRAG, we find a mean $\Delta w = +0.033 \pm 0.026$ and $\Delta w = -0.004 \pm 0.005$ for SALT2.FRAG simulations with SN-only and SN+CMB fits respectively. Similarly, we find a mean $\Delta w = +0.008 \pm 0.026$ and $\Delta w = -0.005 \pm 0.004$ for SALT3.FRAG simulations.

When using all SNe, our $\Delta w$ values are consistent with that of the real data to within at least 2.5$\sigma$, for all simulations except the independent set. We do not compare the independent simulation results to the real data as the analysis conditions are not the same. Instead, the independent simulation results tell us the impact on $w$ of switching SALT models for a \textit{simulation}-based analysis, where the underlying simulated cosmology sample changes along with the light curve fit results. Our $\Delta w$ values are consistent with $\Delta w = 0$ to within at least $1.3\sigma$ for all simulations. This indicates that even though the $\Delta w$ results using the SN data alone are an order-of-magnitude larger than those using SN+CMB constraints, they still yield a SALT-dependent bias that is negligible compared within the statistical constraints of our analysis.

We now move to the results obtained when each surface must only use the SNe that pass light curve fitting cuts with \textit{both} surfaces, i.e., the "Common SNe" rows of Table~\ref{deltaw_table}.
When fitting the same underlying simulation with SALT2.FRAG and SALT3.FRAG, we find a mean $\Delta w = -0.007 \pm 0.007$ and $\Delta w = -0.011 \pm 0.001$ for SALT2.FRAG simulations with SN-only and SN+CMB fits respectively. Similarly, we find a mean $\Delta w = -0.028 \pm 0.006$ and $\Delta w = -0.012 \pm 0.001$ for SALT3.FRAG simulations.

With consistent sampling, the absolute difference in cosmology result between surfaces increases for all simulations (except SALT2.FRAG-generated results fit with SN data only, hereafter SALT2.FRAG-SNONLY). The uncertainty on those differences decreases due to a narrower spread in the results from our 50 simulated samples. For any input simulation except SALT2.FRAG-SNONLY, our results strongly disagree with both the real data and the $\Delta w = 0$ case at tensions of up to $20\sigma$. The difference between the consistently-sampled results from simulations and real data may arise from subtle differences in the application of consistent sampling between the products, or it may be a consequence of the 1D-bias corrections not accounting for this additional selection effect. Usually we expect a cosmology fit with 1D bias corrections to recover our input cosmology ($w=-1$) to $\leq 3\sigma$, as is the case in the ''All SNe" results. However, 5 out of our 8 mean $w$ results for each SALT surface from the consistently-sampled simulations display a $>3\sigma$ tension with the input cosmology, indicating the presence of an uncorrected selection effect.\footnote{e.g., SN-only results from a SALT2.FRAG fit of SALT2.FRAG-generated and consistently sampled SNe yield a mean $w = -1.19 \pm 0.03$.} Modelling this selection effect is beyond the scope of this work, but our preliminary results have shown that this modelling is necessary for any future work that wishes to enforce such consistent sampling in a cosmology analysis. 
\textbf{However, as long as all selection effects are considered in the bias corrections and one does not assume that the samples recovered by SALT2 or SALT3 light curve fitting process will be exactly identical, we conclude that the bias on $w$ from the choice of SALT framework is insignificant relative to the statistical error in a self-consistent analysis.}




\section{Discussion and Conclusions}
\label{sec:discussion}

\subsubsection*{The SALT.FRAG Surfaces}
We have performed a SN cosmology analysis to test for differences in SALT2 and SALT3 SN~Ia light curve models, and find that the choice of model does not bias cosmology results. We have fit SN and cosmological parameters using both real and simulated DES-SN3YR data, and the most advanced SALT2 and SALT3 surfaces released to date (the "FRAG" surfaces, which we present in this paper). The model training sample and calibration was the same for SALT2.FRAG and SALT3.FRAG, so any difference in our fitted results must have arisen directly from the differences in the underlying model frameworks and applications. To facilitate this, we first applied a necessary correction to the SALT2 training process that until now had not been using a fully trained colour law (see Appendix~\ref{app:cl} for details).

The SALT2.FRAG and SALT3.FRAG model components are largely consistent, with smoothed $M_0$ differences below $2.5\%$ over 3500-8500\AA\ (most of the useful model range). The colour laws are also consistent within this range. We have shown that both SALT2.FRAG and SALT3.FRAG are less sensitive to photometric calibration uncertainties than previous releases (Figure~\ref{SALT23_sys}), and published the suite of systematic surfaces needed to calculate these uncertainties. This is an important step towards achieving the science requirements set out by \citet{2018lsst_sciencereqs}.

\subsubsection*{Impact of Choice of SALT.FRAG Surface on Simulations}

Throughout our work, we found evidence that the choice of surface used to generate simulations had a small impact on our results.
The simulated distribution of $x_1$ changes between the two surfaces --- SALT3.FRAG generates more SNe in the secondary low-$x_1$ peak than SALT2.FRAG. This follows from SALT3's updated training procedure that requires $c$ and $x_1$ to be independent.


\subsubsection*{Impact of Choice of SALT.FRAG Surface on Light Curve Fitting}

We have shown that the recovered cosmological parameters depend on the cosmology sample used, which in turn depends on the model used to fit the light curves. SALT3.FRAG is able to successfully fit more SNe~Ia (i.e., more SNe pass cuts) than SALT2.FRAG at low redshifts where there is high signal-to-noise photometry. We found that a portion of this effect originates from overestimated model uncertainties in the SALT3.FRAG surface, which over-fit the data to give a reduced $\chi^2 < 1$ (Figure~\ref{data_deltafitprob}). However, we also found that the SALT3.FRAG model uncertainties correct a sample-dependent effect in SALT2 models, where the fit probability appears to be correlated with SNR. Additionally, we find that SALT3.FRAG tends to cut SNe with higher absolute Hubble residuals than SALT2.FRAG. 

We see comparable results for each surface when comparing the fit light curve parameter values in the data (Figure~\ref{sixpanel_data}).
The trends with redshift between the two surfaces are small, with slopes that are consistent with 0 for all fitted parameters except the 1D bias-corrections.

The fitted uncertainties on the $m_B$ and $\beta c$ light curve parameters reduce when light curve fitting with SALT3.FRAG, regardless of the choice of input model for the simulation. This is also seen in the real data. However, the $\alpha x_1$ uncertainty tends to increase with SALT3.FRAG, leading to comparable $\mu$ uncertainties when fitting both surfaces on the same simulation or data set.

When comparing the fitted values in the simulations with the known input values, we find SALT3.FRAG pull distributions that are wider for $m_B$ and narrower for $x_1, c$ and $\mu$ than the corresponding SALT2.FRAG distributions. This suggests that light curve fitting with SALT3.FRAG yields results that are closer to the true values than those fit with SALT2.FRAG.

\subsubsection*{Impact of Choice of SALT.FRAG Surface on Cosmology}

We find $\Delta w = +0.001$ between SALT2.FRAG and SALT3.FRAG for the real DES-SN3YR data (when combined with a CMB-like prior). We consider this value to be drawn from the distribution of values found from corresponding simulations generated with SALT2.FRAG ($\Delta w = -0.004 \pm 0.005$) or SALT3.FRAG ($\Delta w = -0.005 \pm 0.004$). All of these results are within 1$\sigma$ of our measurement uncertainties on $w$ ($\sim0.06$) and $1.3\sigma$ of the $\Delta w = 0$ (i.e. no bias) case. While we recover larger absolute $\Delta w$ values ($\sim|0.03|$) when only using SN data with no CMB-prior in our cosmology fit, the increased statistical uncertainty means these results are still consistent with the null-bias case.

We also recover large absolute $\Delta w$ values ($\sim|0.03|$) when only using the commonly-fit subset of SNe for each surface, though this type of sampling requirement is unrealistic in SN cosmology analysis and is not recommended due to the complications it introduces in the bias correction process. We therefore report our $w$-bias from the choice of SALT2.FRAG or SALT3.FRAG in a SN+CMB cosmology analysis to be $\Delta w = +0.001 \pm 0.005$, using our real data result and the largest uncertainty calculated from simulations that use the same underlying sample of SNe (\textit{not} the same sample after light curve fitting cuts).

Recent analyses have only quantified the photometric uncertainty inherent in the SALT2 model (i.e. the uncertainty probed by the systematic surfaces described in $\S$~\ref{SALT23_sys}) in their systematic budgets ($\sigma_w^{sys}$ = 0.009 in \citealt{des3yr_systematics}, 0.008 in \citealt{Jones_2018} and 0.023 in \citealt{Jones_2019}). \citet{mosher14, jla} were the last works to calculate and include a more thorough SALT2 systematic uncertainty ($\sigma_w = -0.014 \pm 0.007$), however, they stress that $\sigma_w$ will vary depending on the training sample and cosmology sample. Other estimates of SALT2 model systematic uncertainties compare the cosmology results found with SALT2 versus some other model (e.g., \verb|MLCS2K2|, \citealt{kessler09}) --- but as SALT2 and SALT3 share a large portion of their underlying framework, even these estimates are not a good point of comparison to the $\Delta w$ we find here. \textbf{We therefore provide the \textit{first} estimate of systematic uncertainty arising from the choice of SALT model framework (i.e. SALT2 versus SALT3), $\Delta w$ = +0.001 $\pm$ 0.005.} We stress that our conservative systematic uncertainty estimate does not account for changes in the SALT surface training sample, calibration, or rest-frame wavelength fitting range --- all of which are held constant across SALT2.FRAG and SALT3.FRAG.

\subsubsection*{Conclusions}

We have shown that for models trained on similar inputs, the choice of SALT2 or SALT3 for simulations and light-curve fitting does not bias a cosmology analysis --- though we do recover different samples after light-curve fitting, this is generally accounted for in the bias corrections. We have shown that even when both models are trained on state-of-the-art training samples, SALT3 slightly outperforms SALT2, recovering fitted parameter values that are closer to the true simulated values. These results, along with the ease-of-use of \verb|SALTShaker|, motivate the use of SALT3 over SALT2 for SN cosmology analyses (such as DES-SN5YR) going forward.

SALT3 spans a larger wavelength range than that currently achievable with SALT2, increasing the redshift range over which $z-$ and $Y$-band data can be used to constrain light curves. This improves the quality of SN cosmology results, increases the usable data from already-completed surveys such as DES, and is a step towards being able to separate the intrinsic SN colour from host galaxy extinction. The performance gap between SALT2 and SALT3 will continue to widen with larger training samples, when SALT3 can operate over a wider wavelength range (e.g., \citealt{salt3nir}).

From a model perspective, SALT3 is more physically realistic --- the regularization scheme mitigates a large amount of un-physical negative flux in the model SEDs (Figure~\ref{m0-m1}), and the treatment of uncertainties based on the intrinsic dispersion of the underlying SN~Ia population does not rely on the distribution of the training sample. It is also much more easily maintained and updated than SALT2, owing to its open-source Python-based training framework. It is therefore straightforward to investigate the issue of overestimated model uncertainties raised in this work. We highlight that the problematic correlation of SALT2's fit probability with data sample (likely linked to SNR) is resolved with SALT3.

Future work to understand the systematic uncertainties in the SALT3 model is ongoing. \citet{dai_22} use simulations to consider the impact of variations in the training data. Taylor et al., in prep use simulations to consider the impact of the underlying intrinsic scatter models on the model surface. \citet{jones_22_host} construct a SALT3 model of SN~Ia dependence on host mass. All this, combined with ongoing efforts to observe an increasingly powerful training sample, signal the continued development of an advanced and increasingly-well-understood light curve model for use in the next generation of SN cosmology. SN analyses that make use of SALT should carefully consider the choice of model, training sample, and wavelength fitting range as systematics in their uncertainty budgets.

\section*{Acknowledgements}

This research was supported by an Australian Government Research Training Program (RTP) Scholarship. GT thanks Chris Lidman, Brad Tucker, Tamara Davis, and Dillon Brout for their helpful discussion and comments. A portion of this research was completed during a Lamprecht Residency; GT thanks B., B., and S. Lamprecht for supporting this work. This project used public archival data from the Dark Energy Survey (DES). Funding for the DES Projects has been provided by the U.S. Department of Energy, the U.S. National Science Foundation, the Ministry of Science and Education of Spain, the Science and Technology Facilities Council of the United Kingdom, the Higher Education Funding Council for England, the National Center for Supercomputing Applications at the University of Illinois at Urbana-Champaign, the Kavli Institute of Cosmological Physics at the University of Chicago, the Center for Cosmology and Astro-Particle Physics at the Ohio State University, the Mitchell Institute for Fundamental Physics and Astronomy at Texas A\&M University, Financiadora de Estudos e Projetos, Funda{\c c}{\~a}o Carlos Chagas Filho de Amparo {\`a} Pesquisa do Estado do Rio de Janeiro, Conselho Nacional de Desenvolvimento Cient{\'i}fico e Tecnol{\'o}gico and the Minist{\'e}rio da Ci{\^e}ncia, Tecnologia e Inova{\c c}{\~a}o, the Deutsche Forschungsgemeinschaft, and the Collaborating Institutions in the Dark Energy Survey.
The Collaborating Institutions are Argonne National Laboratory, the University of California at Santa Cruz, the University of Cambridge, Centro de Investigaciones Energ{\'e}ticas, Medioambientales y Tecnol{\'o}gicas-Madrid, the University of Chicago, University College London, the DES-Brazil Consortium, the University of Edinburgh, the Eidgen{\"o}ssische Technische Hochschule (ETH) Z{\"u}rich, Fermi National Accelerator Laboratory, the University of Illinois at Urbana-Champaign, the Institut de Ci{\`e}ncies de l'Espai (IEEC/CSIC), the Institut de F{\'i}sica d'Altes Energies, Lawrence Berkeley National Laboratory, the Ludwig-Maximilians Universit{\"a}t M{\"u}nchen and the associated Excellence Cluster Universe, the University of Michigan, the National Optical Astronomy Observatory, the University of Nottingham, The Ohio State University, the OzDES Membership Consortium, the University of Pennsylvania, the University of Portsmouth, SLAC National Accelerator Laboratory, Stanford University, the University of Sussex, and Texas A\&M University. 
Based in part on observations at Cerro Tololo Inter-American Observatory, National Optical Astronomy Observatory, which is operated by the Association of Universities for Research in Astronomy (AURA) under a cooperative agreement with the National Science Foundation.
This work was completed in part with resources provided by the University of Chicago Research Computing Center.

\section*{Data Availability}
The surfaces described in this article are publicly available at 10.5281/zenodo.4001177 and 10.5281/zenodo.7400436. The data underlying this article are available at \url{https://des.ncsa.illinois.edu/releases/sn}.




\bibliographystyle{mnras}
\bibliography{refs} 

\appendix

\section{Impact of Colour Law Update}
\label{app:cl}

\begin{figure}
  \begin{center}
   \includegraphics[width=\columnwidth]{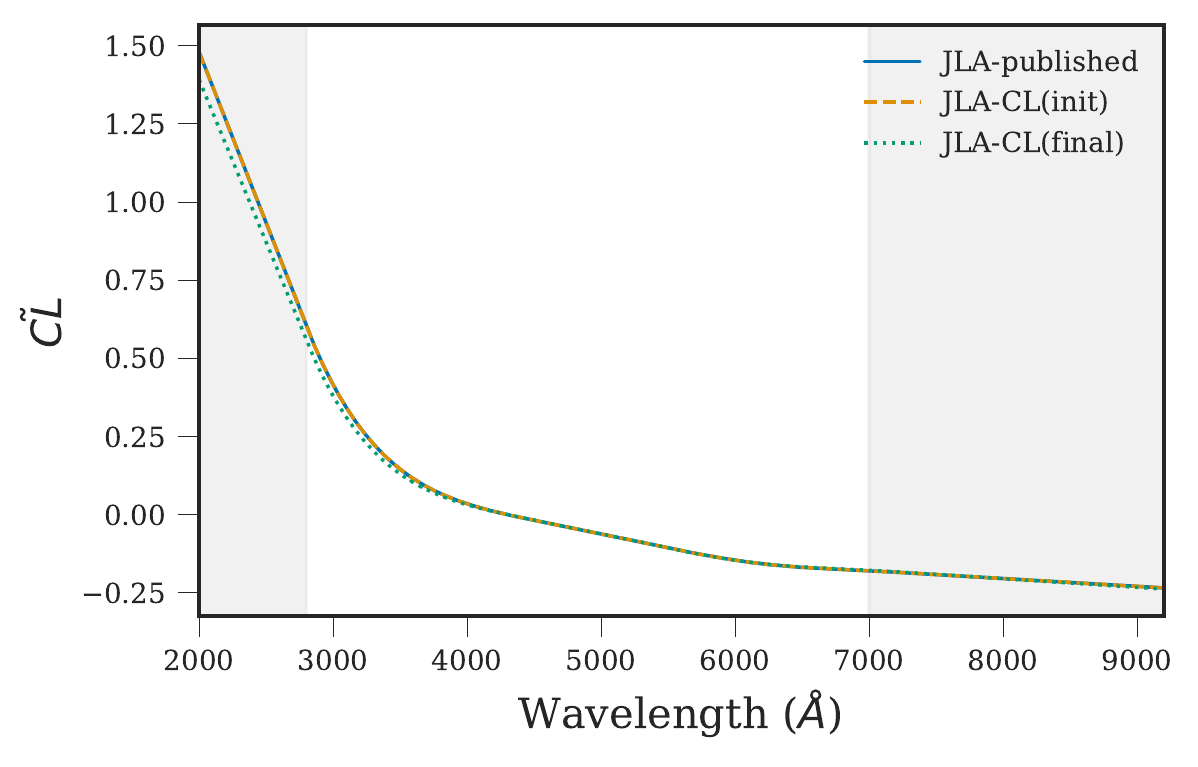}
  \end{center}
  \caption{The published JLA colour law (\textit{solid blue)} versus the initial (\textit{dashed orange}) and final (\textit{dotted green}) JLA colour laws, for a $c=-0.1$ SN~Ia, where $\tilde{CL} \equiv c CL(\lambda)$. This shows that the initial colour law was published with the SALT2.JLA surface.}
  \label{JLA_cl}
\end{figure}

As noted in $\S$~\ref{section:SALT2-training}, the colour law previously used by SALT2.JLA (and consequent surfaces trained under the same \verb|snpca| scheme) appears to use an incorrect file. An initial fit of the colour law is used, rather than the final trained colour law (as prescribed in e.g., \citealt{mosher14}). We correct the training scheme for the SALT2 surface presented in this work, and in this section briefly discuss the impact of the updated colour law training scheme on a DES-SN3YR-like analysis.

\begin{figure}
  \begin{center}
   \includegraphics[width=\columnwidth]{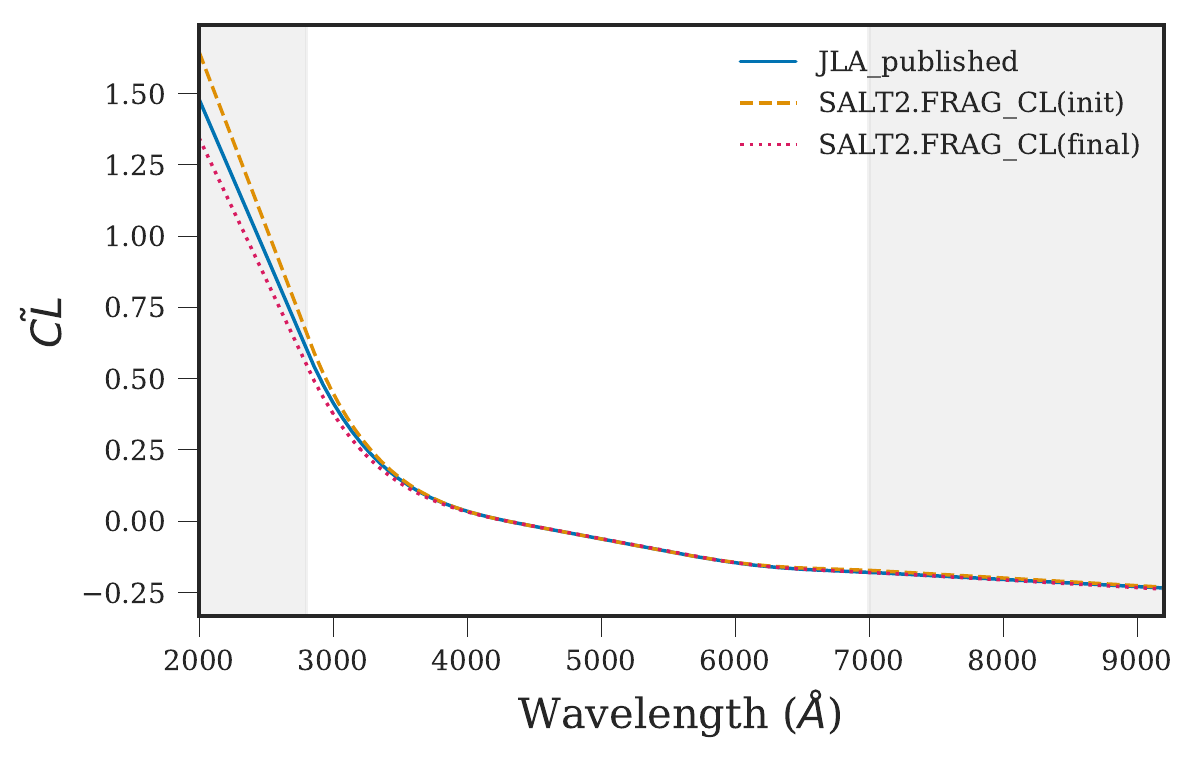}
  \end{center}
  \caption{The published JLA colour law (\textit{solid blue)} versus the initial (\textit{dashed orange}) and final (\textit{dotted pink}) SALT2.FRAG colour laws, for a $c=-0.1$ SN~Ia where $\tilde{CL} \equiv c CL(\lambda)$. This shows this larger impact of the colour law update when applied to the \citetalias{salt3} training sample used for SALT2.FRAG.}
  \label{T22_cl_init}
\end{figure}

Figure~\ref{JLA_cl} plots the published JLA colour law, the "initial" JLA colour law, and  the ``final" (corrected) JLA colour law. As the final colour law file is generated during the last step of SALT2 training, the other components ($M_0$ and $M_1$) are unaffected by this update. The differences are most significant below 4000\AA, which should mitigate the impact of this change on published cosmology results which use SALT2.JLA.
Figure~\ref{T22_cl_init} shows the impact of the updated training scheme on the colour law of the SALT2.FRAG surface presented in this work. \textbf{We saw a redshift-dependent change in SN colour for the DES-SN3YR analysis when fitting with the SALT2.FRAG initial versus final colour laws, highlighting the importance of this update for SALT2 surfaces going forward.} 

\begin{figure*}
  \begin{center}
   \includegraphics[width=2\columnwidth]{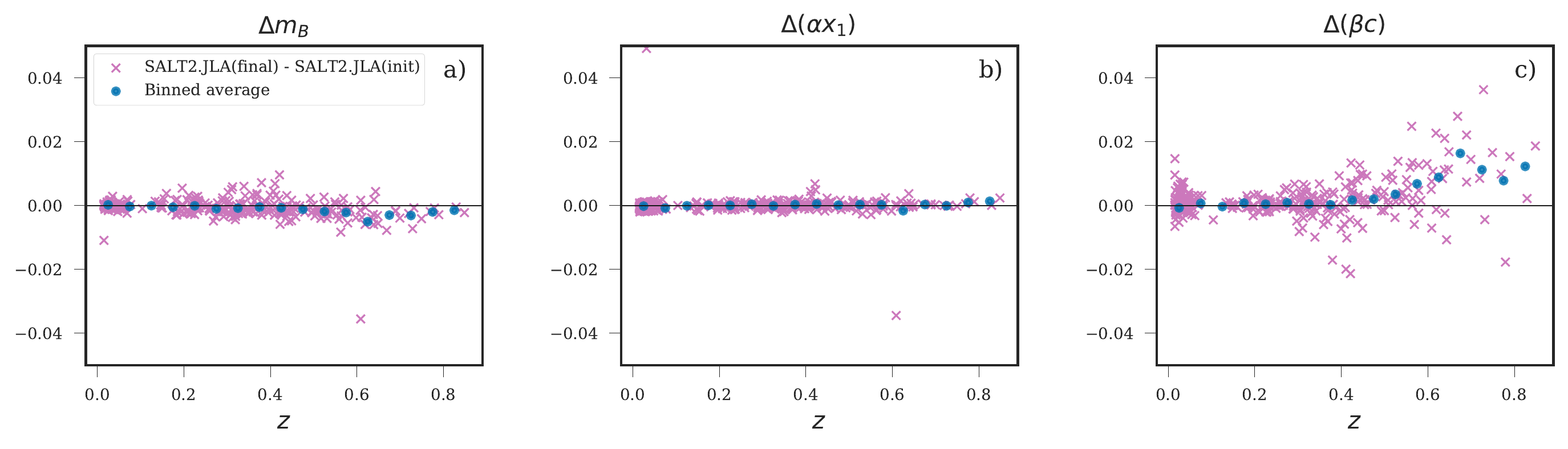}
  \end{center}
  \caption{Differences in fitted light curve parameters (for DES-SN3YR data) between SALT2.JLA (initial colour law) and SALT2.JLA (final colour law), plotted against redshift. No bias corrections are applied to the parameters. Two outlier SN are not shown.}
  \label{cl_lcparams}
\end{figure*}

\begin{figure}
  \begin{center}
   \includegraphics[width=\columnwidth]{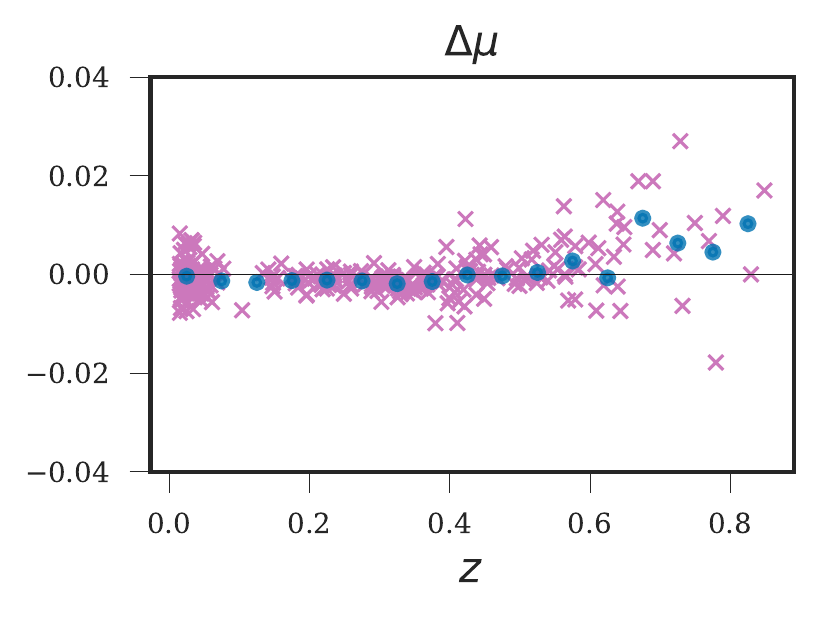}
  \end{center}
  \caption{Differences in fitted distance moduli (for DES-SN3YR data) between SALT2.JLA (initial colour law) and SALT2.JLA (final colour law), plotted against redshift. No bias corrections are applied. Two outlier SN are not shown.}
  \label{cl_muparams}
\end{figure}

We estimate the impact of this change on published cosmology results with SALT2.JLA by fitting the DES-SN3YR sample (including the accompanying low-$z$ sample) using both the initial and final JLA colour laws. The results (shown in Figure~\ref{cl_lcparams}) do not include bias corrections and so may not fully capture all changes. We see a redshift-dependent change in SN colour (Figure~\ref{cl_lcparams}c) when moving from the initial to final JLA colour law, with the strongest effects at higher redshifts (where there are fewer SNe per redshift bin). The overall impact on $\mu$ (calculated without bias corrections) is shown in Figure~\ref{cl_muparams}, which closely follows the trend in SN colour from Figure~\ref{cl_lcparams}c. A rudimentary analysis with \verb|wfit| finds $\Delta w = -0.0027$, which is an order of magnitude smaller than the total $\sigma_w^{stat+sys}$ used in the published DES-SN3YR analysis \citep{des3yr_systematics}. 

\section{Common Model Definitions for SALT2 and SALT3}
\label{common_model_definitions}

To constrain degeneracies in the SALT model, some definitions are applied during the training. Definitions that are unique to either SALT2 or SALT3 were described in Section \ref{section:23_differences}; the following definitions are common to both:

\begin{enumerate}[label=\arabic*)]
    \item The rest-frame synthetic $B$-band flux of the $M_0$ component at peak is fixed such that $m_B^{peak} = 10.5$ when $x_0 = 1$.
    \item The rest-frame synthetic $B$-band flux of the $M_1$ component at peak is defined to be 0.
    \item The distribution of $x_1$ in the training sample is defined to have mean~=~0 and a standard deviation = 1.
    \item The distribution of $c$ in the training sample is defined to have mean~=~0.
    \item The colour law is defined such that $CL(4300$\AA$)=0$ and $CL(5430$\AA$)=-1$, corresponding to central wavelengths for $B$ and $V$ bandpasses.
\end{enumerate}

These conditions mean that fitted $x_1$ and $c$ values for a cosmology sample are defined relative to the training sample of the SALT model used for fitting. One should therefore be careful when comparing fitted parameters from models with different training samples.

\begin{table*}
\renewcommand\thempfootnote{\arabic{mpfootnote}}
\begin{minipage}{\textwidth}
  \renewcommand{\arraystretch}{1.5}
  \centering
  \caption{Parent populations for DES and Low-Z SNe~Ia when fit with SALT2.FRAG and SALT3.FRAG surfaces. Following \citet{Popovic_2021} we fit asymmetric Gaussians to the populations for each parameter, with the exception of the low-z $x_1$ distribution which instead uses a double-peaked Gaussian.}
    \label{parentpop_table}
\begin{tabular}{|c|l|c|c|}
\hline
\multicolumn{1}{|l|}{} &  & \textbf{SALT2.FRAG} & \textbf{SALT3.FRAG} \\ \hline
\multirow{7}{*}{\textbf{Low-Z}} & \textit{Mean $c$} & $-0.052 \pm 0.021$ & $-0.048 \pm 0.022$ \\ \cline{2-4} 
 & \textit{Std. $c$ (left)} & $0.020 \pm 0.014$ & $0.030 \pm 0.014$ \\ \cline{2-4} 
 & \textit{Std. $c$ (right)} & $0.141 \pm 0.033$ & $0.145 \pm 0.039$ \\ \cline{2-4} 
 & \textit{Weight $x_1$ (1)\footnote{The low-$z$ stretch distribution is double-peaked; the weight parameter scales the respective Gaussian. See Equation~5 of \citet{Popovic_2021}.}} & $2.057 \pm 1.010$ & $2.99 \pm 1.02$ \\ \cline{2-4} 
 & \textit{Mean $x_1$ (1)} & $-1.658 \pm 0.105$ & $-1.545 \pm 0.284$ \\ \cline{2-4} 
 & \textit{Std. $x_1$ (1)} & $0.409 \pm 0.119$ & $0.471 \pm 0.347$ \\ \cline{2-4} 
 & \textit{Weight $x_1$ (2)} & $3.286 \pm 1.197$ &   $2.992 \pm 1.190$ \\ \cline{2-4} 
 & \textit{Mean $x_1$ (2)} & $0.429 \pm 0.062$ & $0.423 \pm 0.113$ \\ \cline{2-4} 
 & \textit{Std. $x_1$ (2)} & $0.405 \pm 0.692$ & $0.509 \pm 0.110$ \\ \hline
\multirow{6}{*}{\textbf{DES}} & \textit{Mean $c$} & $-0.077 \pm 0.018$ & $-0.068 \pm 0.021$ \\ \cline{2-4} 
 & \textit{Std. $c$ (left)} & $0.020 \pm 0.013$ & $0.031 \pm 0.014$ \\ \cline{2-4} 
 & \textit{Std. $c$ (right)} & $0.126 \pm 0.015$ & $0.126 \pm 0.015$ \\ \cline{2-4} 
 & \textit{Mean $x_1$} & $0.933 \pm 0.272$ & $0.865 \pm 0.287$ \\ \cline{2-4} 
 & \textit{Std. $x_1$ (left)} & $1.631 \pm 0.217$ & $1.569 \pm 0.215$ \\ \cline{2-4} 
 & \textit{Std. $x_1$ (right)} & $0.308 \pm 0.196$ & $0.329 \pm 0.207$ \\ \hline
\end{tabular}
\end{minipage}
\end{table*}

\section{Parent Populations for SALT2.FRAG and SALT3.FRAG}
\label{app:parentpops}

The underlying SNe~Ia populations (as they appear in nature) for the surfaces used in this work, given our analysis methods, are presented in Table~\ref{parentpop_table}. These were calculated using the Parent Populations program, available at \url{github.com/bap37/ParentPops}.

\bsp	
\label{lastpage}
\end{document}